\documentclass[10pt,journal,cspaper,compsoc]{IEEEtran}
\usepackage{url}
\usepackage{multirow}
\usepackage[ruled,vlined]{algorithm2e} 
\usepackage{amsfonts}
\usepackage{rotating}

\ifCLASSOPTIONcompsoc
 
\else

\fi

\ifCLASSINFOpdf

\else

\fi

\newtheorem{definition}{Definition}

\begin{document}
%
\title{ACTIDS: An Active Strategy For Detecting And Localizing Network Attacks}

\author{Eitan Menahem, Gabi Nakibly, Yuval Elovici
\IEEEcompsocitemizethanks{\IEEEcompsocthanksitem E. Menahem, and Y. Elovici are with the Telekom Innovation Laboratories and information System Engineering Department, Ben-Gurion University, Be'er Sheva, 84105, Israel and G. Nakibly is with Rafael -- Advanced Defense Systems Ltd., Haifa, Israel \protect\\
Email:\{eitanme,elovici\}@post.bgu.ac.il, gabin@rafael.co.il}
\thanks{}
}


\IEEEcompsoctitleabstractindextext{%
\begin{abstract}
In this work we investigate a new approach for detecting attacks which aim to degrade the network's Quality of Service (QoS). To this end, a new network-based intrusion detection system (NIDS) is proposed. Most contemporary NIDSs take a passive approach by solely monitoring the network's production traffic. This paper explores a complementary approach in which distributed agents actively send out periodic probes. The probes are continuously monitored to detect anomalous behavior of the network. The proposed approach takes away much of the variability of the network's production traffic that makes it so difficult to classify. This enables the NIDS to detect more subtle attacks which would not be detected using the passive approach alone. Furthermore, the active probing approach allows the NIDS to be effectively trained using only examples of the network's normal states, hence enabling an effective detection of zero-day attacks. Using realistic experiments, we show that an NIDS which also leverages the active approach is considerably more effective in detecting attacks which aim to degrade the network's QoS when compared to an NIDS which relies solely on the passive approach. Lastly, we show that the false positives rate remains very low even in the face of Byzantine faults. 
\end{abstract}

}

\maketitle

\IEEEdisplaynotcompsoctitleabstractindextext

%
\IEEEpeerreviewmaketitle

\section{Introduction}
Network intrusion detection systems (NIDS) are key in the security architecture of many organizations. A typical NIDS inspects traffic flowing into, out of, or inside the target network while attempting to isolate a malicious activity. Broadly speaking, an NIDS tries to identify two types of malicious activities. The first type includes attacks carried over network traffic that target end nodes. One such example is a worm that aims to infect and take control over end nodes. Another example is a reconnaissance activity that scans active IP addresses or opens ports at some end nodes. The second type of attack targets the network itself. Attacks of this type include exhaustion of a link's bandwidth by overwhelming it with traffic as well as unauthorized modification of a network's routing process. The ultimate goal of such attacks is to degrade the QoS provided by the network to its users.

Traditionally, NIDSs are broadly classified based on the style of detection they use. Some systems rely on a precise description of the malicious activity, i.e., knowledge-based (signature-based) detection. Other systems rely on statistical modeling of the network's normal state and regard significant deviations from this state as attacks, i.e., anomaly-based detection.

The current paper addresses the problem of detecting QoS degrading attacks which target the network itself, and further focuses on attack classes which have not yet been seen before, i.e., zero-day attacks. In order detect new attacks classes, an IDS should not rely on the description of known attacks (e.g., DNS poisoning or OSPF attacks), their specific features (e.g., increased number of failed TCP connections) or even on monitoring known attacks surfaces (e.g., DNS cache or routing table), since a new attack class might exploit the vulnerabilities of previously unexplored attack surfaces, which might have significantly different features from those of previous attacks. Consequently, the knowledge-based IDS approach is not suitable for the task at hand. Hence, in the rest of the paper we focus on the anomaly-based detection approach.

Most anomaly-based NIDS take a passive approach. They rely exclusively on monitoring the network's production traffic and extracting the relevant features that indicate the progression of an attack. However, experience has shown that the immense variability of network traffic is a major stumbling block of the NIDS~\cite{PaxonSommer10}.  Such variability is demonstrated in many of the network's traffic features and consequently makes them very difficult to predict over short time scales (seconds to hours) and furthermore presents difficulties in detecting anomalies generated by the network attacks. In addition, a passive NIDS needs to process all the production traffic flowing through it. It has been shown that anomaly-based NIDS are unable to run at line rate in the network core~\cite{Lakhina05}. In order to mitigate this problem, an NIDS may be deployed closer to the network edge where its visibility of the entire network traffic is reduced. Alternatively, packet flow sampling may be used, however, that can further degrade the detection accuracy~\cite{Brauckhoff06}.

Our work takes a somewhat different approach and aims to complement the pitfalls of the passive approach. We propose to produce artificial probing traffic exchanged between various agents over the network. General QoS features are then extracted from the probing traffic, rather than from the production traffic. Lastly, anomalies are identified based on those features. The rationale of this approach is as follows. First, the probing traffic is much more predictable than the production traffic and therefore an abrupt change in one of the features gleaned from that traffic is most likely sourced at an anomaly of the network rather than a legitimate shift of traffic. This approach takes away much of the variability of network traffic which makes it so difficult to predict. Secondly, the probing traffic is treated by the network as regular production traffic and therefore probing traffic will be affected by network attacks as much as the production traffic is. Thus, the detection potential is not diminished.

Note that a fundamental property to this approach is that it detects anomalies produced by the attack \emph{effects} on the network, such as changes in traffic delays or packet losses. This allows us to detect any attack, including  zero-day attacks, as long as it affects the network's QoS parameters. In particular, we do not attempt to monitor the by-products effects of specific types of attacks, such as the increased number of failed TCP connections, increased traffic volume, shifts in port or IP distribution or changes to routing tables and DNS caches.
This will allow us to identify a wide range of attacks which target the network, regardless of their nature or the techniques they employ. Examples of such attacks include but are not limited to: 1) subversion of the routing process in the network; 2) poisoning of basic network services, such as the DNS; and 3) overloading of the network's links or routers. Overall, our NIDS architecture aims to detect every attack that has some adverse effect on the service provided to users by the network. 

Another key feature of our work is the use of one-class learning. Our NIDS is trained only on a normal network state. This means that network attack records and specifications are not required, further creating  substantial operational benefits. The key assumption of this strategy is that it is possible to obtain a complete picture of almost all normal states in the network and thereby reliably infer that any other state is anomalous. This assumption is known as the Closed World assumption~\cite{weka}. In general, the assumption is believed to be impractical for most real-life problems since it is impossible to cover all normal states. Nonetheless, our work focuses on the network'’s effects and the QoS it provides to its users. This QoS is generally regarded as stable and predictable to the extent that many network operators have commercial obligations to it under service level agreements (SLAs). Therefore, our working assumption is that we can cover all normal network states with a high degree of confidence. 
Furthermore, it has been shown~\cite{Barford09} that a deviation of network performance for operational purposes can be successfully detected using anomaly detection techniques. 

\subsection{Applicability}
The NIDS architecture proposed in this work aims to identify and locate attacks that have an adverse effect on the network and service provided to its users. That is, our aim is not to find attacks on the network's end hosts that do not affect the network itself, such as the silent spread of a worm in the network, a slow port scan or a buffer-overflow that affects only the victim end host. Therefore, it is important to note that we do \emph{not} aim to propose an alternative to existing detection approaches, but rather to complement them where they are less effective. 

Our assumption is that the attacker's goal is to do one of the following: (a) inhibit or reduce the service provided by the network to its users (e.g., DoS attacks) or (b) modify services provided by the network (e.g., routing process or other basic services such as DNS). As mentioned above, we aim to detect all such attacks regardless of the specific techniques they employ or the protocol they use over the IP protocol. 

Our approach may detect rare benign events that affect the network's QoS, such as a flash crowd, as anomalies. However, in section \ref{sec:LinkFailures}, we show that by using a simple method, our proposed NIDS can learn to ignore such benign anomalies with a very small cost of intrusion detection delay.\\

Our NIDS architecture identifies the location of the attack's effect, not the attack's traffic or its source. The rationale of this approach is based on the fact that the detection of the true source of a real-world attack demands elaborate forensic work conducted by a human expert. The findings or alerts of any NIDS are only the starting point to detecting the true source of the attack. This is especially true in the case of the common multi-stage attacks~\cite{Clark10}. 

The proposed NIDS architecture is focused on intra-domain network settings, such as enterprise or ISP networks. In such a setting, the network operator has the ability to deploy the proposed distributed architecture. In addition, the operator has full knowledge of the network topology and the routes taken between every pair of end nodes. This knowledge is essential in order to deploy our proposed architecture efficiently.

\subsection{One-Class Learning}
One-class learning is a special case of machine learning whose main goal is to differentiate examples of the class of interest from all other examples. The one-class classifier is trained from a training-set containing only the instances of that class. 
In network security, one-class classifiers most frequently train on the normal state of the network, i.e., they model the network's normal state. There are two main features that make one-class classifiers attractive for network security. First, they do not require any labeled attack instances, which are very difficult and costly to attain and second, they can identify zero-day attacks, which are often the attacker's weapon of choice.

\subsection{One-Class Ensemble Learning}

One of the contributions of this work is the application of meta learning ensemble to NIDS. The main idea behind this ensemble methodology is to weigh several individual classifiers (ensemble members) and to combine them via another classifier (combiner) to obtain a classifier that outperforms the ensemble members. Theoretically, an ensemble can benefit from any independent base-classifier that performs even slightly better than a random one (see the Condorcet Jury Theorem and \cite{Schapire90}). Indeed, previous studies in supervised ensemble learning (e.g., \cite{MenahemRE09}) show that combining classification models can produce a better classifier in terms of prediction accuracy. 

In contrast to the research on supervised learning, progress in the theory of combining one-class classifiers has been limited. According to \cite{GiacintoPRR08}, up until 2008, this research field was relatively new and had not been thoroughly explored. In particular, in the setup of diverse ensemble members, only two combining methods were considered for one-class problems: the \emph{fix-rule} \cite{Tax01, JuszczakD04} and meta-learning \cite{DBLP:journals/corr/abs-1112-5246} ensembles.

Meta-learning is a process of learning from basic classifiers (ensemble members); the inputs of the meta-learner are the outputs of the ensemble-member classifiers. The goal of meta-learning ensembles is to train a meta-model (meta-classifier) that will combine the ensemble members' predictions into a single prediction. In order to create such ensembles, both the ensemble members and the meta-classifier need to be trained. Since the meta-classifier training requires already trained ensemble members, these must be trained first. The ensemble members are then used to produce outputs (classifications) from which the meta-level dataset (meta-dataset) is created. This dataset will be used for training the meta-classifier. The basic building blocks of meta-learning are the meta-features, which are measured properties of the ensemble members, e.g., an ensemble members' predictions. A vector of meta-features comprises a meta-instance, i.e., meta-instance $\equiv <f^{meta}_{1},...,f^{meta}_{k}>$. A collection of meta-instances comprises the meta-training-set upon which the meta-classifier is trained. 

\subsection*{Paper Outline}
The rest of this paper is organized as follows. In Section \ref{sec:related} we present related work. In Section \ref{sec:method}, we introduce the proposed NIDS method. 
We then discuss the experimental setup and the network attacks used in the evaluation in Section \ref{sec:methodologies} and present the results in Section \ref{sec:results}. Finally, we close with conclusions in Section \ref{sec:conclusions}.

\section{Related Works}
\label{sec:related}

In the past two decades, many anomaly-based NIDS schemes have been proposed. The common trend among these systems is that they model the network's normal state by analyzing aspects of the production traffic, e.g., network flows \cite{
Lee01realtime,SalemVG10,DBLP:journals/tdsc/GuptaNR10}, packet payload information \cite{
DBLP:journals/tdsc/WangZS04,DBLP:journals/tdsc/HwangCCQ07, HubballiBN10}, network event analysis \cite{Sommer03, OlivainG05, BenaliBGC10} or management information base (MIB) aggregated information \cite{Bao09, Jou00designand, NassarSF10}. 

By depending solely on the production traffic (or its aggregations), of which they have no control, these NIDS take a passive approach and should hence be labeled ``\emph{passive-NIDS}.'' The passive approach has many advantages like high data availability and simplicity. However, many drawbacks also exist. Sommer et al.\ \cite{PaxonSommer10} list five attributes of the network security domain that usually prohibit (passive) anomaly-based NIDS from being implemented or adopted in real networks. They argue that detecting attacks is a very different and more difficult task than the classic task of machine-learning. Their list of five attributes is as follows: 
First, ``real-life'' errors bear a very high cost in reality. 
Second, usually only a very limited training data is available. 
Third, there is a difference between the anomaly detector output and the operational meaning.  
Fourth, it is difficult to train a sound model of the network's normal state because of the huge variance in the training data. 
Fifth, machine-learning NIDS cannot be evaluated in conditions sufficiently close to reality and therefore their benchmark results do not reflect their eventual performance on real networks.

In contrast to passive-NIDS, the active-NIDS approach, by which the NIDS relies on self-traffic incurred by probe sending, can overcome some of the aforementioned deficiencies.
A fundamental property of the active-NIDS is that during the normal network operation, its probe features, such as \emph{hop\_count} and \emph{average\_round\_trip\_time}, are characterized by very low variability. This means that probes' features are very predictable, thus can be modeled easily and effectively by training classifiers on limited size datasets, as we show in Section \ref{sec:results}. While classification errors are not easy to avoid completely, maintaining the very costly false alarms at zero, while at the same time detecting every significant attack, is indeed possible, but at the cost of a detection time delay. In this study we demonstrate such a technique and measure its detection time delay (we refer to this as Time-to-Detect, denoted $TtD$). 

Byzantine faults, which might occur due to unexpected network failures or topology changes, should be detected by the NIDS if presenting a significant change from the network's normal state. We argue that these benign events should be addressed by the network operator and should not be counted as false alarms of the NIDS. In appendix \ref{sec:LinkFailures}, we demonstrate a method for reducing the detection of benign anomalies while still preserving the detection rate.
The active-NIDS has full knowledge of the paths its probes travel during training time; therefore, when an anomaly occurs, it can localize the affected network devices by identifying the probes with anomalous features. Such a technique is given in Section \ref{Localization}. The anomaly localization feature provides the security specialist and the network operator with valuable information. Although this does not close the gap between anomaly detection output and operational meaning, it is a step closer to this goal.
 
Up until now, the active probing approach has been practiced mainly for network monitoring, fault localization and network diagnosis, e.g., \cite{DBLP:journals/ton/BejeranoR06, DBLP:conf/infocom/ChengQMQB10, DYSWIS08}.
Our survey found only two related studies. The first, by Barbhuiya et al. \cite{BarbhuiyaBN11}, presents a genuine NIDS, whereas the research by Barford et al. \cite{Barford09} is actually related to network QoS assurance, though is quite easy to translate for network-security purposes.

The active-NIDS studied by Barbhuiya et al.\ \cite{BarbhuiyaBN11} sends address resolution protocol (ARP) probes to detect ARP spoofing attacks. The authors devised a formal model of the protocol state transitions with which they later detect anomalies. 
The main drawbacks of their method are that it is protocol-specific, namely, it is not sufficiently general to detect other classes of attacks and it protects the hosts only in the local network.

Barford et al. \cite{Barford09} also take the active probing approach but with the intentions of identifying network QoS anomalies and localizing anomalous links. Their ultimate goal is to detect and localize QoS anomalies that might occur on \emph{any} existing path between \emph{any} two measurement nodes. To do so, they must cover the entire network by a probing scheme. To avoid producing and analyzing $O(N^2)$ probes in each time unit, they limit the monitoring task to only $k$ selected (with probability) paths in each time unit; since they assume that the anomalies are persistent, they are certain to detect them after finite time periods. Furthermore, they perform the anomaly detection and localization task in two stages; first an anomaly detection stage, followed by the anomaly localization stage. 

While this approach works for QoS assurance, it is not entirely suitable for the network security domain, since network security attacks that cause link failures are intentional, malicious, and might be limited in time, while in operational networks, link failures tend to occur sporadically and are persistent.

Moreover, the two-phase detection and localization method has been criticized \cite{SalhiLC10} as producing sub-optimal anomaly detection and localization performance. An additional limitation of Barford et al.'s method is that they put no bound on the length of their probes, and by not doing so, they are usually required to send elevated numbers of probes to localize an anomalous link, in contrast to a short probing scheme, such as the one described in the present study.

\section{The Proposed Method}
\label{sec:method}
The new NIDS scheme, ACTIDS (Active Intrusion Detection System), is mainly characterized by five features: (1) dispatching of probe packets to generate network traffic, from which features are extracted; (2) hierarchical anomaly detection architecture, i.e., the network is partitioned into smaller disjointed sub-networks (a single autonomous anomaly classifier (local anomaly classifier) is linked with each network sub-network); (3) anomaly classifiers are trained on examples of the normal network state (i.e., \emph{one-class}) so attack examples are not required during training; (4) anomaly detection and localization are performed simultaneously; and (5) in combining the predictions of the local anomaly classifiers, a state-of-the-art meta-learning ensemble is used.

\subsection{Architecture}
ACTIDS includes three main components: Agents, Sensors, and a Central Detection Server (CDS). The hierarchy is as follows. Each Sensor gathers information from multiple Agents, which are the elementary parts, and the CDS combines information from multiple Sensors. These three components are described in depth below. 
	
\emph{\textbf{Agent}} - lightweight software responsible for transmitting probe packets and extracting QoS statistics from incoming probe packets. The Agents are installed at selected hosts. The probe packets are standard IP packets which carry special protocol values in order to be distinguished from other packets at the network layer. Agents are linked to at least one Sensor.
	
\begin{figure}[ht]
	\centering
		\includegraphics[viewport=0 120 800 600, scale =0.2]{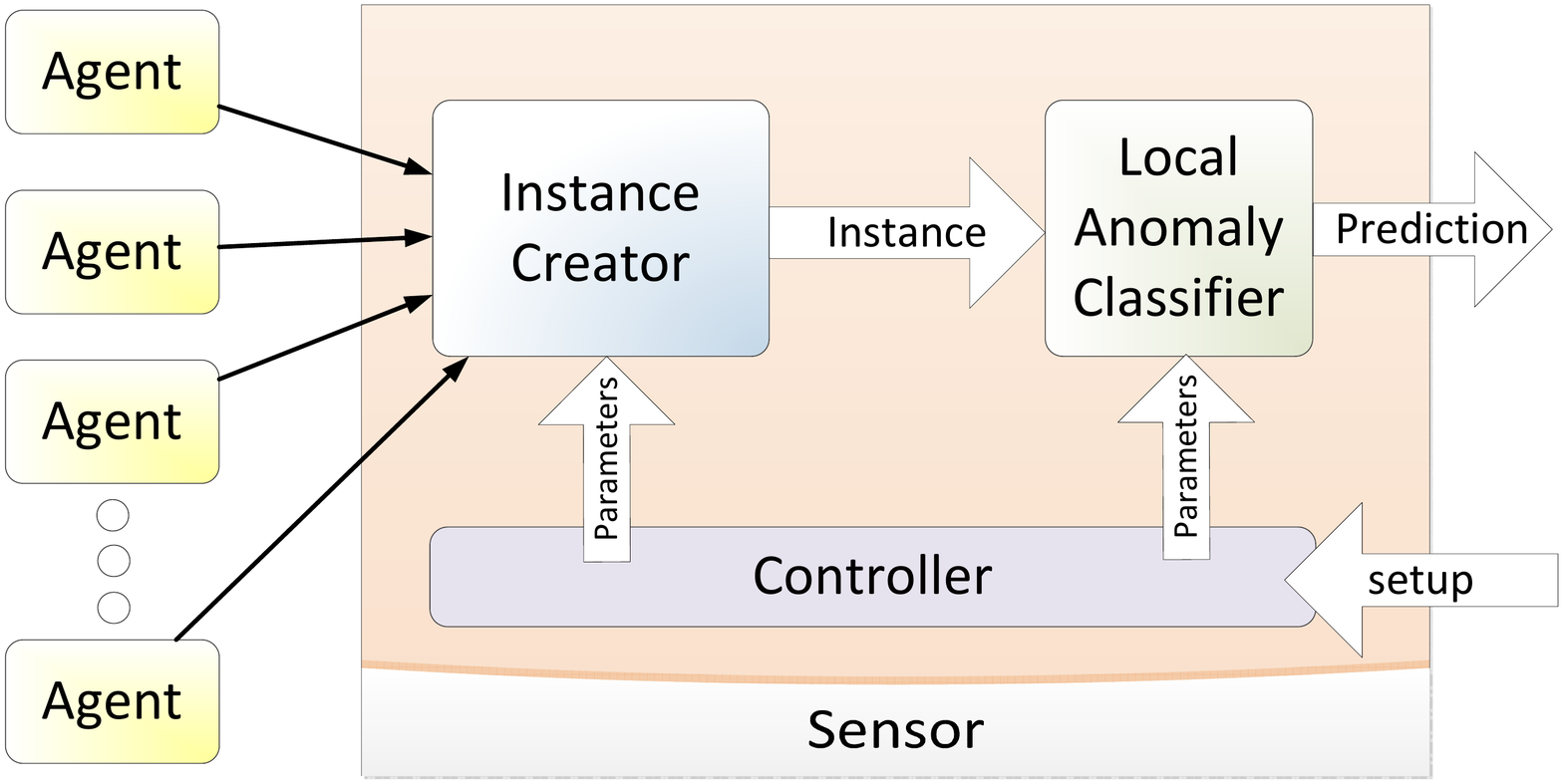} 
	\caption{The Sensor's conceptual architecture}
	\label{fig:Sensor}
\end{figure}

\emph{\textbf{Sensor}} - an anomaly detector overseeing a limited-size connected network section. During the setup phase (see \ref{sec:setup}), the network is partitioned into disjointed sections, each of which is assigned to a single Sensor. 
The Sensor, as can be seen in Figure \ref{fig:Sensor}, is made up of an Instance Creator module and a machine-learning anomaly classifier and is bound to Agents installed at its network section.  The Instance Creator module continuously aggregates data from the bound Agents, and once every pre-defined time period, it produces an instance that is then classified by the anomaly classifier. The classification or anomaly score is a numerical value in the range of [0,1] which represents the likelihood of local anomaly and is outputted to the CDS, where it is combined with other Sensors' classifications.

\emph{\textbf{Central Detection Server (CDS)}} - produces the global anomaly score, which indicates the security condition of the entire network. The CDS plays a role in all three states of the proposed NIDS. In the setup phase, the CDS has two functions. First, it determines the network partition, according to the \emph{partition algorithm}; each partition holds a single Sensor. The CDS's second function during setup is to determine the probing scheme. During the training phase, the CDS's anomaly classifier is trained. Lastly, in the detection phase, the CDS produce a global anomaly score by combining anomaly scores produced by the Sensors.

\begin{figure}[ht]
	\centering
		\includegraphics[viewport=0 0 800 600, scale =0.25]{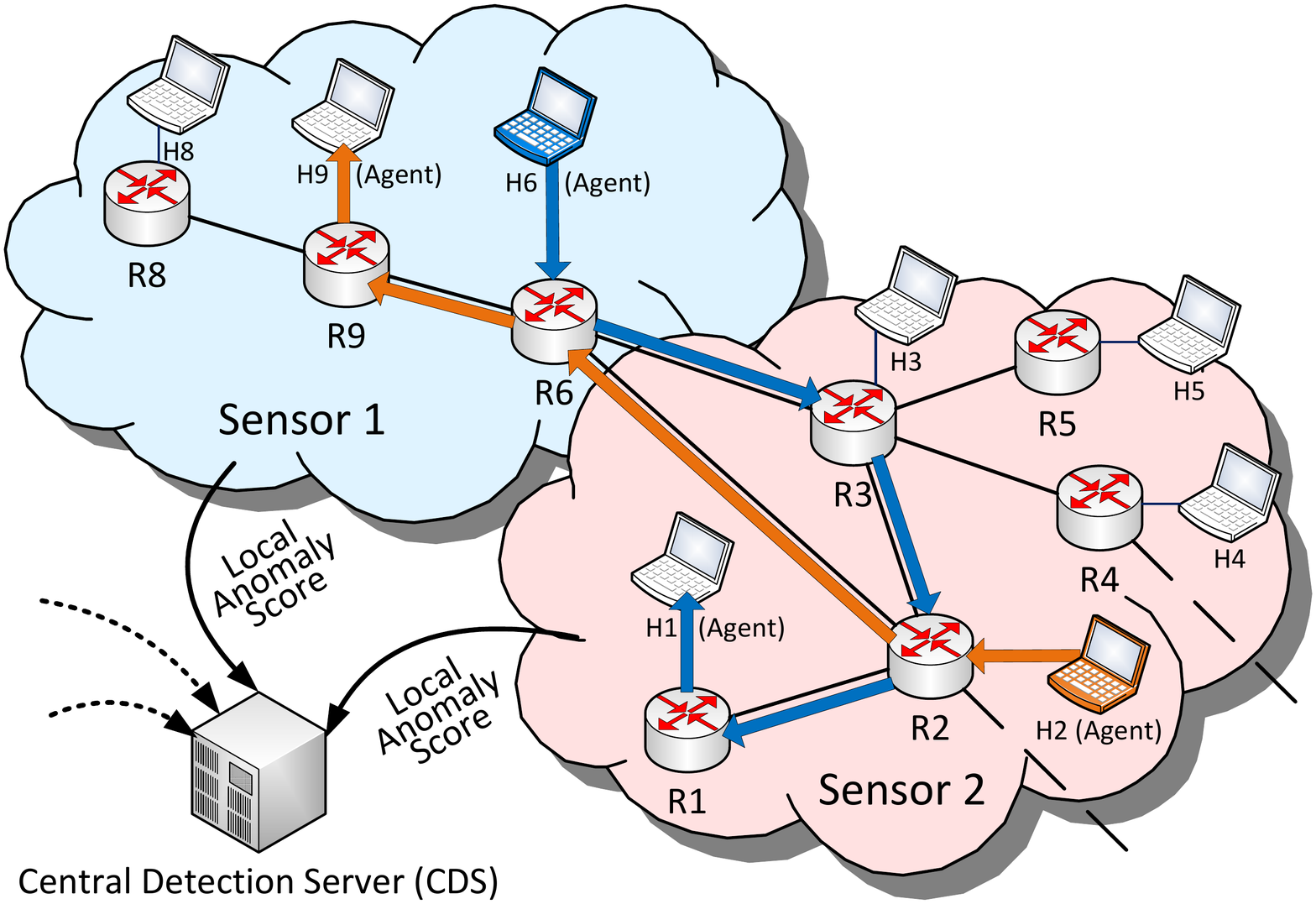} 
	
	\caption{Example of two probes, sent across two Sensors, which then send their corresponding local anomaly score to the CDS}
	\label{fig:ProbingScheme}
	\vspace{-3mm}
\end{figure}

ACTIDS has a three stage life cycle: setup, training, and prediction. All the aforementioned stages are fully automatic and require very limited user intervention. We now describe each of these stages.
\subsection{Setup Phase}
\label{sec:setup}
In the setup stage two pre-processes of ACTIDS are executed; the first determines the Sensors' partition while the second determines the probing scheme. 

\subsubsection{Local Anomaly Sensors}
During the setup phase, the network is logically partitioned into sub-networks. A local anomaly detector (Sensor) is attached to each sub-network. Algorithm \ref{alg:partitioning} establishes a set of disjoint Sensors, each of which contains up to \emph{max\_number\_of\_routers} routers.

\vspace{2mm}
\begin{definition}
\label{def:Probe}
(Probe). \emph{Probe is a special media that is sent between two Agents.}
\end{definition}
\vspace{2mm}
The Probes are used for extracting crude network features upon which anomalies are detected. Probes make a roundtrip, starting at the source Agent, traveling to the destination Agent at one of the adjacent Sensors, and then returning to the source. The roundtrip is necessary in order to avoid router time synchronization, which may be difficult to attain in real-life networks.
A Probe can be made of any type of media which is appropriate for sending over the network, e.g., IP packet, ICMP packet or any application packet of choice. The source and destination hosts relate to different Sensors.
\begin{definition}
\label{def:ProbePath}
(Probe-Path). Probe-Path is a set of connected edges $(v,e)$ which follow the shortest path between the Probe's source host and its destination host and back to the source host. 
\end{definition}
\begin{definition}
\label{def:Probe Cycle Time}
(Probe re-sending). Let $T_{prs}$ denote the constant time interval between subsequent Probe re-sending.
\end{definition}

\subsubsection{Probe Dispatching Mechanism}
The \emph{Probing Mechanism} produces a list of Probes. Later, during the training and detection phases (the second and third stages of the proposed NIDS's life cycle, respectively), this Probe list is sent to enable Agents to measure statistics regarding the network's QoS. During the training and detecting stages, the Probes are re-sent once every fixed (configurable) period of time. A procedure for an automatic generation of a probing scheme is presented in Algorithm \ref{alg:probeScheme}. We call this algorithm ``the probing algorithm.'' The probing algorithm is the second and final algorithm related to the setup phase. The inputs for this algorithm are the Sensors' partition, a list of routers, and the Probe length, i.e., \emph{probesLength}. 
The probing scheme algorithm follows three guidelines: 

\begin{enumerate}

	\item To increase the likelihood of detecting anomalies, Probe-Paths should cover as many of the network's edges as possible.
	\item To reduce both Probe traffic and learning complexity, use as few Probes as possible.
	\item To enable accurate detection of the anomaly source, Probe-Paths should be short, i.e., 4 or 6 edges.
\end{enumerate}

These guidelines can be translated into an optimization problem in which we aim to cover all the network's edges with a minimal set of Probe-Paths over a bounded number of edges. A special case of this problem, in which no bound on the Probe-Paths' length exists, has been shown to be an instance of the NP-hard \emph{Minimum Set Cover} problem~\cite{DBLP:journals/ton/BejeranoR06}. 
However, covering all the network's edges is very likely to generate too much Probe traffic and increase the learning dimensionality, thus making the learning task too complex. 

Our probing scheme uses only on the order of $O(|v|)$ Probe-Paths while attempting to cover most of the network devices. This is because each router sends, at most, a single Probe, whose path length is bound by the $pathLength$ constant, as presented in Algorithm \ref{alg:probeScheme}.
 
Such a probing scheme will generate sufficient information to detect attacks whose effects reach beyond the boundaries of a single Sensor. 
As the network attack affects a greater section of the network, and is therefore regarded as more severe from the perspective of the defender, the probability of the effect being captured by at least one Probe increases.

\begin{algorithm}[ht]
\DontPrintSemicolon

\KwIn{$Network Topology$ : network's topology}
\KwIn{\emph{max\_routers} : the maximal number of routers in each Sensor}
\KwOut{SensorsMap : a list of Sensors}
\SetKwFunction{ParseTopology}{ParseTopology}\SetKwFunction{Sort}{Sort}\SetKwFunction{deleteFirst}{deleteFirst}\SetKwFunction{CalculateDegree}{CalculateDegree}\SetKwFunction{createEmptySensor}{createEmptySensor}\SetKwFunction{DeleteNeighbor}{DeleteNeighbor}
$Routers\leftarrow$\ParseTopology{$Network Topology$}
$SensorsMap\leftarrow\emptyset$\;
\SetAlgoVlined
\ForEach{$r$ in Router}{$Degree\leftarrow$ \CalculateDegree{$r$}\;}
\Sort{$Routers$} by $Degree$;\

\While{$|Routers|>0$}{
	$currentSensor\leftarrow$\createEmptySensor()\;
	$kernel \leftarrow$\deleteFirst($Routers$);\
	
	Add $Kernel$ To $currentSensor$\;
	\While{$|currentSensor| \leq max\_routers$}{
		$nr\leftarrow$\DeleteNeighbor($kernel$,$Routers$)\;
		Add $nr$ To $currentSensor$\;
	}
	Add $currentSensor$ To $SensorsMap$
}

\caption{Partition the network into disjointed Sensors}
\label{alg:partitioning}
\end{algorithm}

The \emph{DeleteNighbor(r,Routers)} routine extracts the neighbor router of $k$ with the highest rank and deletes it from router group $Routers$.

\begin{algorithm}[ht]
\begin{minipage}{0.9\linewidth}
\KwIn{$Routers Map$, $Sensors Map$, \emph{probesLength}}
\KwOut{list of Probes}
\DontPrintSemicolon
\SetKwFunction{RoutersAtDistance}{RoutersAtDistance}
\SetKwFunction{OnSameSensor}{OnSameSensor}
\SetKwFunction{CreateEmptyProbe}{CreateEmptyProbe}
\SetKwFunction{GetSensor}{GetSensor}
\SetKwFunction{GetAdjacentSensor}{GetAdjacentSensor} 
$pl \leftarrow probesLength$\;
$ProbeList\leftarrow\emptyset$\;
\ForEach{$r$ in $RouterMap$}{
	$srcS \leftarrow$ \GetSensor($r$,$Sensors\-Map$)
	$Candidates \leftarrow $\RoutersAtDistance($r$,$pl$)\;
	\ForEach{$c$ in $Candidates$}{
	$destS \leftarrow$ \GetSensor($c$,$Sensors\-Map$)\;
	\If{Not($srcS$==$destS$) AND \\
		  ExistProbeBetween($srcS$,$destS$)=false}{
		$src \leftarrow r$\;
		$dest \leftarrow c$\;
		$ProbeList$+=New Probe($src$,$dest$)\;
		end foreach\;
		}
	}
}
\caption{Probe-Scheme Generation}
\label{alg:probeScheme}
		\end{minipage} 
\end{algorithm}
  
\subsection{Training Phase}
In this stage all the machine-learning models, i.e., the local anomaly detectors attached to the Sensors and the global anomaly classifier in the CDS, are trained. At the end of this stage, the proposed NIDS is ready for the detection phase. The training process comprises three distinct tasks: producing the training datasets, training the Sensors' classifiers, and training the global anomaly classifier. We now discuss each of these tasks.  
\begin{definition}
\label{def:Re-Classification Time}
(Classification Time). Let $T_{cl}$ denote the constant time interval between behavior classifications of subsequent  networks.
\end{definition}
Prior to training the machine-learning-based classifiers, a set of appropriate training data is produced. This is achieved in several stages. First, the NIDS begins sending Probes across the network according to the pre-acquired Probe-Scheme. The Probe traffic is added to the network's normal chaotic traffic. As the Probes are re-sent once every $T_{prs}$ during the entire training phase, they incur the traffic from which the Agents collect raw statistics. When an Agent receives a Probe, it extracts three raw features; \emph{hop count}, \emph{travel time}, and \emph{num\_lost\_probes} 
 and delivers them to the Sensor to which it is attached. The Sensor aggregates the received statistics until, once per $T_{cl}$ ($T_{cl}>>T_{prs}$), it computes the average and the variance of the data accumulated for the first two raw features, computes the \emph{lost\_probes percentage}, and constructs an instance. The instance is then added to the Sensor's private training-set. Next, the statistics at the Sensors are nullified and the process repeats until the end of the training phase, by which time all the training-sets contain the same number of instances. Table \ref{tab:SensorSTrainingSet} presents the Sensors' training-set structure.

Next ,when the Sensors' training-sets generation is done, the Sensors' classifiers are trained, each on its own private training-set. Lastly , ACTIDS uses the TUPSO ensemble scheme \cite{DBLP:journals/corr/abs-1112-5246} to train the CDS's classifier, which is responsible for combining the Sensors' local anomaly score during the prediction phase.



\begin{table}[b]
\scriptsize
	\centering
		\begin{tabular}
		{@{} p{5mm}|p{7mm}|p{7mm}|p{12mm}|p{12mm}|@{\hspace{0.1mm}}p{9mm}@{\hspace{0.1mm}}|@{}p{0.1mm}@{}|@{}p{0.1mm}@{}|@{}p{0.1mm}@{}|@{}p{0.1mm}@{}|@{}p{0.1mm}@{}}
		\hline
		\multicolumn{1}{r|}{} & \multicolumn{5}{c|}{\footnotesize Probe: $R8 \longrightarrow R2$} & \multicolumn{5}{c@{}}{...} \\
		\hline
		Time Unit & Avg. Delay & Var. Delay & Avg. Hop Count & Var. Hop Count &  \% Lost Probes& \multicolumn{1}{@{}c@{}|}{...} & \multicolumn{1}{@{}c@{}|}{...} & 
		\multicolumn{1}{@{}c@{}|}{...} & \multicolumn{1}{@{}c@{}|}{...} & \multicolumn{1}{@{}c@{}}{...}\\ 
		\hline
		1 & \multicolumn{1}{c|}{ $a_{1,1}$} & \multicolumn{1}{c|}{ $a_{1,2}$} & \multicolumn{1}{c|}{ $a_{1,3}$} & \multicolumn{1}{c|}{ $a_{1,4}$} & \multicolumn{1}{c|}{ $a_{1,5}$} &  	\multicolumn{1}{@{}c@{}|}{...}  &  \multicolumn{1}{@{}c@{}|}{...}  &  \multicolumn{1}{@{}c@{}|}{...}  & \multicolumn{1}{@{}c@{}|}{...} & \multicolumn{1}{@{}c@{}}{...}  \\
		\hline
		2 & \multicolumn{1}{c|}{ $a_{2,1}$} & \multicolumn{1}{c|}{ $a_{2,2}$} & \multicolumn{1}{c|}{ $a_{2,3}$} & \multicolumn{1}{c|}{ $a_{2,4}$} & \multicolumn{1}{c|}{ $a_{2,5}$} &   \multicolumn{1}{@{}c@{}|}{...}  &  \multicolumn{1}{@{}c@{}|}{...}  &  \multicolumn{1}{@{}c@{}|}{...} &  \multicolumn{1}{@{}c@{}|}{...} & \multicolumn{1}{@{}c@{}}{...} \\
		\hline
		... & \multicolumn{1}{c|}{...} & \multicolumn{1}{c|}{...} & \multicolumn{1}{c|}{...} & \multicolumn{1}{c|}{...} & \multicolumn{1}{c|}{...}&  \multicolumn{1}{@{}c@{}|}{...}  &  \multicolumn{1}{@{}c@{}|}{...}  &  \multicolumn{1}{@{}c@{}|}{...}  & \multicolumn{1}{@{}c@{}|}{...}  & \multicolumn{1}{@{}c@{}}{...}  \\
		\hline
		\end{tabular}
	\caption{The Sensor's training-set structure. $a_{t,i}$ denotes the value of feature $i$ at time unit $t$}
	\label{tab:SensorSTrainingSet}
	\vspace{-7mm}
\end{table}
%
%
%
%
%
Let  $P^{m}_t= <p^{m}_{t,1},...,p^{m}_{t,n}>$  denote the vector containing the Sensors' predictions, $p^{m}_{1},...,p^{m}_{n}$ for time unit $t$, where $n$ is the number of Sensors. TUPSO generates a special train-set for the CDS's combiner classifier by applying $k$ dedicated aggregate functions (described in \cite{DBLP:journals/corr/abs-1112-5246}), $f(\cdot)$, to $P^{m}$.  Each such aggregation represents a single feature (i.e., column) in the combiner's train-set.
Table \ref{tab:GlobalAnomalyDetectorTrainingSet} illustrates the general structure of the CDS's training-set. $ma_{t,i}$ denotes the value of feature $i$ at time unit $t$.

\begin{table}[t]
\footnotesize
	\centering
		\begin{tabular}{@{}c@{\hspace{.5mm}}|c@{\hspace{1mm}}|@{\hspace{1mm}}c@{\hspace{1mm}}|@{\hspace{0.5mm}}c@{\hspace{0.5mm}}|@{\hspace{1mm}}c@{\hspace{1mm}}|@{\hspace{1mm}}c|c@{}}
		\hline
		Time Unit & $f_{1}\left(P^{m}_t\right)$ & $f_{2}\left(P^{m}_t\right)$ & $f_{3}\left(P^{m}_t\right)$ & ... & $f_{6}\left(P^{m}_t\right)$ & $f_{7}\left(P^{m}_t\right)$ \\
		\hline
		1 & $ma_{1,1}$ & $ma_{1,2}$ & $ma_{1,3}$ & ... & $ma_{1,6}$ & $ma_{1,7}$\\
		2 & $ma_{2,1}$ & $ma_{2,2}$ & $ma_{2,3}$ & ... & $ma_{2,6}$ & $ma_{2,7}$\\
		... & ...  & ... & ... & ... & ... & ...\\
		\hline
		\end{tabular}
	\caption{The training-set structure of the CDS's combiner}
	\label{tab:GlobalAnomalyDetectorTrainingSet}
	\vspace{-3mm}
\end{table}

\subsection{Prediction Phase}
The prediction phase represents the “on-line” state of the proposed NIDS. 
To generate the special traffic required by the proposed NIDS, the Agents continuously send Probes once every $T_{prs}$, according to the Probe-Scheme. Upon receiving the Probes, the Agents extract the same raw features as they did in the training phase. Each Agent adds the features it has extracted to the attached Sensor, where they are accumulated. Once every $T_{cl}$, each Sensor averages the accumulated values and generates a local instance. These instances are then classified by the Sensors' classifiers and their statistics are reset. Lastly, the predictions are combined by TUPSO, which outputs the ensemble prediction for time $t$, denoted as $EP_{t}$.

\subsubsection{NIDS Output}
\label{NIDSOUTPUT}
The NIDS's ensemble output for time $t$, $EP_{t}$, is capable of classifying a snapshot of the network's behavior as it reflects the combined anomaly scores of the local Sensors. However, in situations where the NIDS Sensors are very sensitive, they may incur an erratic global prediction and thus induce false alarms (i.e., normal network behavior mistakenly classified as anomalous). 
To refine its classification quality, the presented NIDS uses the exponentially weighted moving average technique (EWMA) which produces smoother and conservative anomaly scores and, as a result reduces, the potential for false alarms. This, as we show in Section \ref{sec:results}, is one of the factors responsible for incurring extremely low false-positive rates while at the same time allowing for a high detection rate. 
In a nutshell, the EWMA technique accumulates anomalous behavior “evidence” over time and, where there is sufficient evidence, will output an “anomaly” classification. The NIDS output, $NAS\in[0,1]$, is computed as follows: 
\[NAS_{t}=NAS_{t-1}+\xi*\Delta(t,p)\] where $\Delta(t,p)$ is the difference between the weighted average of two time-series of the same length, $p/2$. The first series begins at $t-p$ and the second at $t-p/2$:
\[\Delta(t,p)=\frac{\sum_{i=0}^{\frac{p}{2}-1}{(\frac{1}{\frac{p}{2}-i}*(EP_{t-\frac{p}{2}+i}-EP_{t-p+i}))}}{\sum_{i=1}^{\frac{p}{2}}{(\frac{1}{i})}} \]

Variable $t$ is a discrete time point with granularity $t_{cl}$, and $p$ is the sum length of the time-series taken into consideration; $\xi\in\mathbb{R}^{>0}$ is the amplitude assigned to changes in the network behavior; $NAS_{t}$ represents the likelihood of a network-wide anomaly in time $t$.

\subsubsection{Localization Strategy}
\label{Localization}
Identifying the location of an attack's effect is a task that the network operator has to deal with often. Finding the anomaly location can be highly valuable, particularly while the attack is on-going; this allows for measures which can be deployed to mitigate the attack. Identifying an attack’s effect is also important in the post-attack period, as it may provide the forensic expert with helpful facts regarding the attacker's sources, such as attack propagation. In this section we explain how ACTIDS makes use of the hierarchical structure in order to pinpoint the location of attack's effect. 

To identify the anomaly location, ACTIDS relies heavily on the independent ability of each Sensor to indicate the existence of a local anomaly. 
Once a Sensor detects anomalous behavior, it sends the CDS a computed (local) anomaly score along with a list of network components in the region of which it suspects has anomalous behavior. This list is produced by extracting the network components affecting the anomalous features. Next, the CDS processes the lists and produces a shortlist of suspected anomalous components.

\begin{figure}
	\centering
		\includegraphics[scale=0.41]{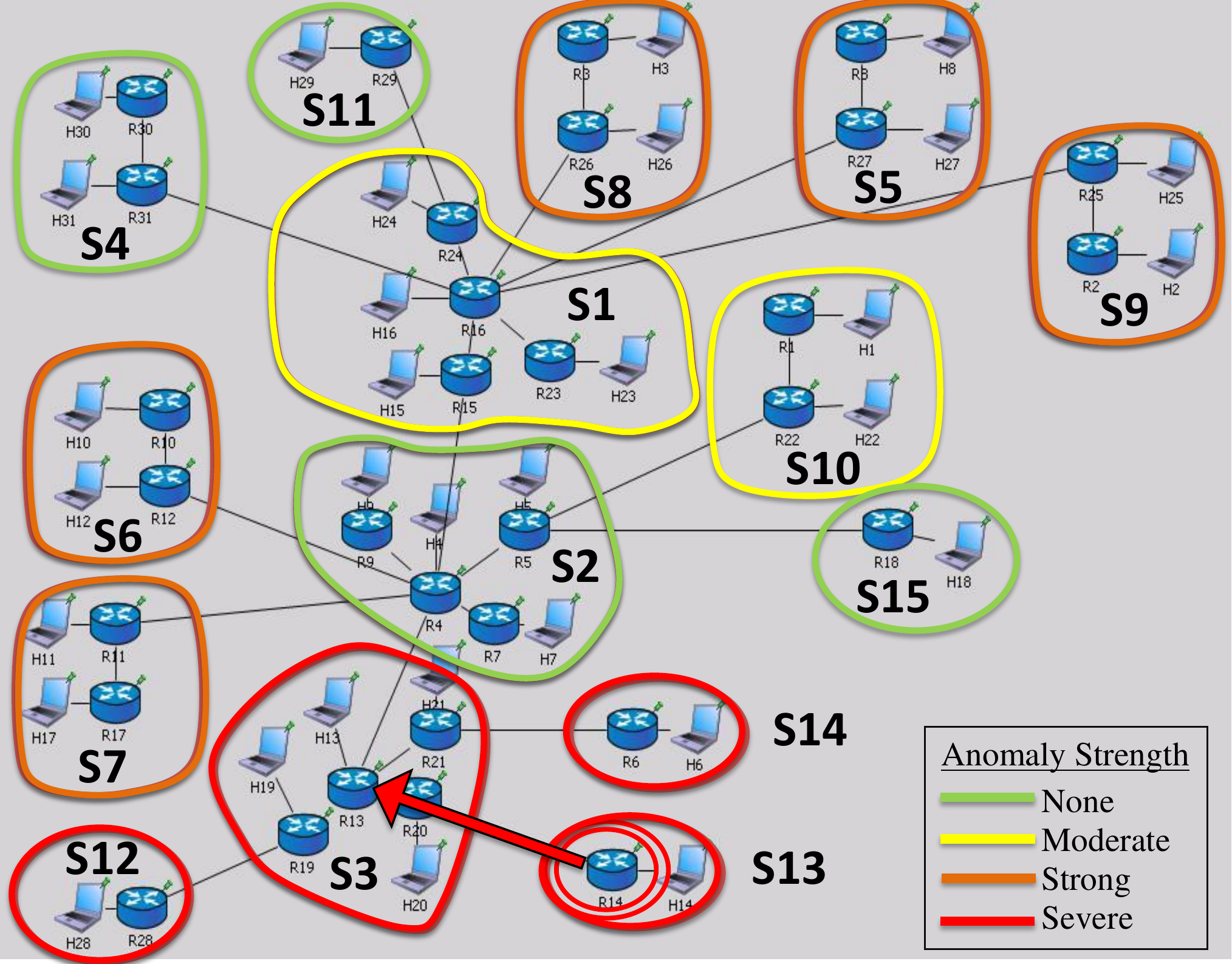} 
	\caption{Sensor scheme deployed on top of AS4755 network topology. In this example R13 is under attack. This attack incurs a network-wide anomaly detected in remote Sensors, such as S5 and S9}
	\label{fig:1755}
	\vspace{-5mm}
\end{figure}

Figure \ref{fig:1755} exemplifies this for the case of R14 attacking R13. Assuming that, because of this attack, the values of three Probe features become anomalous
: $R4 \rightarrow R20$, $R6\rightarrow R13$ and $R28 \rightarrow R13$. This causes Sensors 2, 12, and 14 detect an anomaly. Each of these three Sensors will therefore compile a list of routers that have affected the Probes' values, i.e., the routers that exist in the path of the anomalous Probes. In our example, the reported lists contain: \{R4, \emph{R13}, R20\}, \{R28, R19, \emph{R13}\}, and \{R6, R21, \emph{R13}\} from Sensors S2, S12, and S14, respectively.
Next, to find the anomalous network component(s), the CDS fuses these lists using a very simple rule: 
Let the maximal appearance of the network component be denoted as $maxApp$. It declares any network component in the reported lists that appears \emph{more} than $maxApp/2$ as anomalous. 
In this example, R13 appears three times, while the remaining routers appear only once; therefore, only R13 is declared as anomalous. 
The above process is executed for every test instance; thus, on every classification, ACTIDS produces a list of network components suspected of being anomalous. 
The accuracy of this method depends on four factors: the number of Sensors, the length of the Probes, the anomaly detection accuracy of the Sensors, and lastly, the ability of the Sensor to discriminate between anomalous and normal features. 
Both the Sensors' size limit and the Probes' length were taken into consideration in the design of ACTIDS, hence the limit of routers per Sensor and Probe-Path length were considered as well.

\section{Experimental Setup}
\label{sec:methodologies}
In this section, we specify the conditions in which the proposed NIDS was investigated. First, we illustrate the network simulation used for executing the NIDS. Then, we describe the machine-learning algorithms that participated in the evaluation. Lastly, we present the performance metrics needed to measure the effectiveness of the NIDS.

The NIDS's machine-learning modules were implemented within the WEKA machine-learning framework \cite{weka}. To evaluate the performance of the proposed NIDS, we used the OMNeT++~ \cite{VargaH08} simulation framework, which allows a network to be simulated with full IP stacks. We used real router-level ISP topologies, as mapped by the RocketFuel project~\cite{SpringMW02}. Table~\ref{tab:Rocketfuel} lists the ISP topologies used, which include between ten and several hundred routers. The traffic between the routers was generated based on the well-known gravity model \cite{Kowalski95modellingtraffic}. In this model each router is randomly given a weight, and the average traffic volume between two routers is proportional to the product of their weights. The actual pattern of the generated traffic was based on the “self-similar” model, which is widely believed to be the model that best fits Internet traffic~\cite{self-similar}. 

\begin{table}[hbt]
	\centering
	\footnotesize
		\begin{tabular}{p{35pt} @{\hspace{1.5mm}}@{\hspace{1.5mm}}p{80pt} @{\hspace{1.5mm}}p{35pt}p{35pt}}
			\hline
			Topology & Name &  Routers & Links \\
			\hline
			AS1755 & Ebone (Europe) & 163 & 300 \\
			\hline
			AS3257 & Tiscali (Europe) & 276 & 400 \\
			\hline
			AS3356 & Level3 (U.S.) & 624 & 5,300 \\
			\hline
			AS4755 & VSNL (India) & 11 & 12 \\
			\hline			
			
		\end{tabular}
	\caption{Properties of the autonomous systems used in the evaluation. Source: RocketFuel}
	\label{tab:Rocketfuel}
	\vspace{-5mm}
\end{table}

The simulation was executed for 4500 time units ($TU$); the simulation was run for 1020 $TU$ before the NIDS was executed to ensure the OSPF converged, then it began to send probes at a rate of 25 per $TU$ for the duration of the experiment, which was set to 3500 $TU$. The classification period, $t_{cl}$, was defined as 1 $TU$ so that overall, the datasets (training- and test-sets) contained 3480 instances. Since the Sensors contained differing numbers of data sources (i.e., number of Probes), their datasets varied by the number of features.
Throughout the evaluation, we used \(max\_routers=4\) and \(probesLength=4\) and a classification threshold $\Theta=0.75$ over $NAS$, which was sufficiently high so as to produce zero false positives while maximizing the true-positive rate.

\subsection{Attack Scenarios}
\label{sec:AttackScenarios}

In general, there are two ways for an attacker to harm the QoS of a network. The first way is to change the routes the normal production traffic takes. This way, the attacker may direct traffic through a narrow link or lengthen its route in order to increase the delay and the packet-loss the traffic experiences. The second way to harm the QoS of a network is to induce extra production traffic that would normally not have been generated. One example of this is a reflected SYN attack in which the attacker sends many spoofed SYN packets to various end hosts who respond with SYN-ACK packets. In this example, the extra traffic is induced by the attacker itself (SYN packets), as well as the victim end hosts (SYN-ACK packets). By inducing a high volume of traffic an attacker may exhaust the bandwidth of one or more links in the network.

We evaluated our NIDS only against attack scenarios of the first type, as it is considered more subtle and harder to detect. There are various ways for an attacker to implement such attacks. For example, he may advertise false routing advertisements in order to change the routing tables, he may poison a DNS cache in order to change the destination IP address of the traffic; or he may launch any application-specific impersonation attack in order to divert traffic to a false end host. From the network's point of view, and consequently from our NIDS's point of view, all such attacks have a similar effect, namely, traffic diversion. 

\begin{figure}[h]
	\centering
		\includegraphics[scale=0.22]{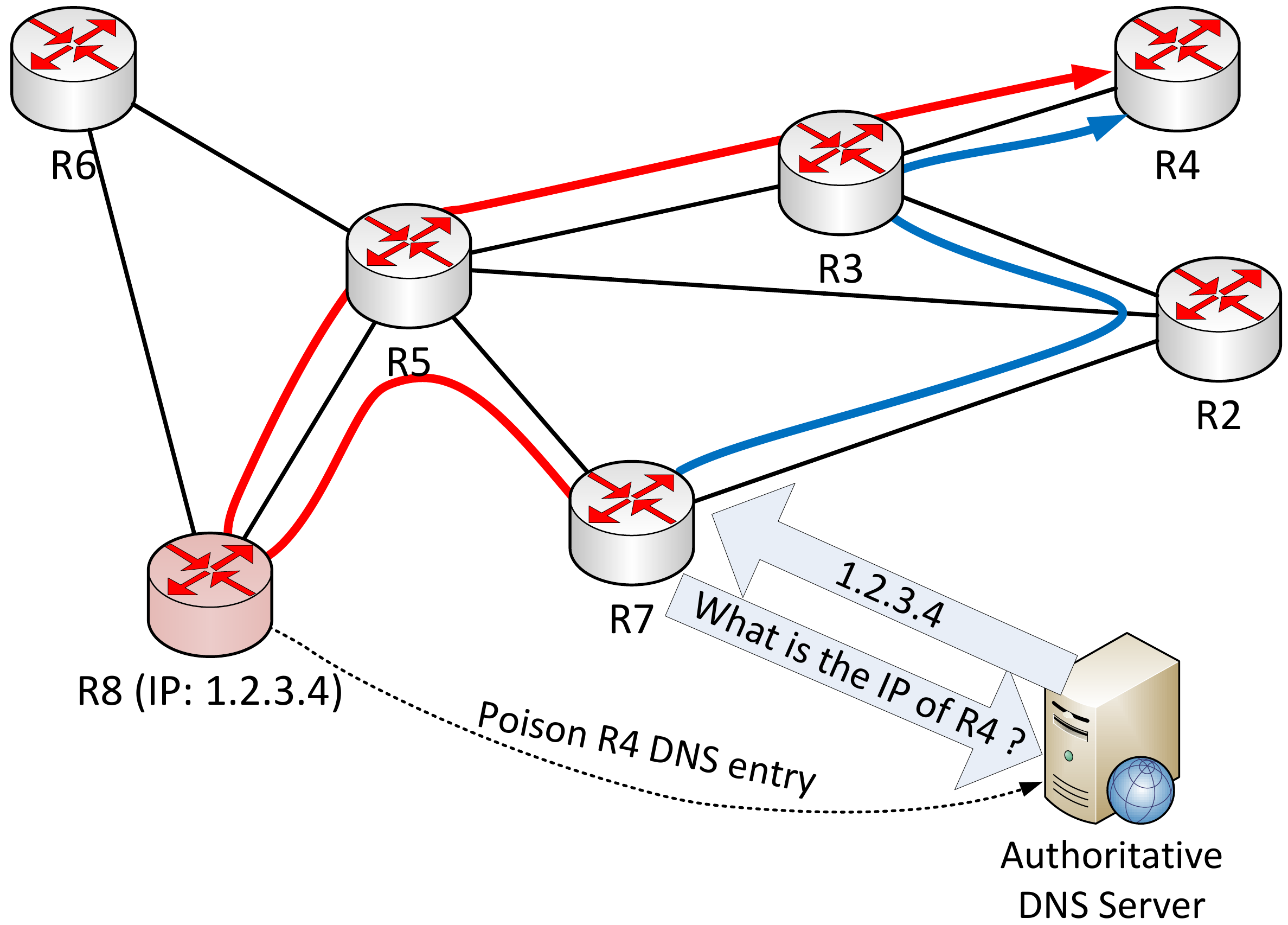}
	\caption{An example of the authoritative DNS server poisoning attack, in which the affect of a single poisoned Authoritative DNS server entry is illustrated.  The blue lines indicate the optimal routes, and the red lines indicate the routes that the packet would take due to the attack}
	\vspace{-3mm}
	\label{fig:Attack-Scenaio_Authoritative_DNS}
\end{figure}

To evaluate our NIDS's effectiveness against traffic diversion attacks we executed two OPSF attacks, i.e., partially disconnecting and link-weight distortion attacks, and two DNS attack variants, i.e., DNS cache poisoning and authoritative DNS server poisoning (shown in Figure \ref{fig:Attack-Scenaio_Authoritative_DNS}). The description of these four attack variants is given in \cite{DiversionAttacks}. These attacks are \emph{currently} the most flexible and powerful way to achieve traffic diversion. Please note that ACTIDS is not a protocol-specific solution, i.e., it is not limited to the abovementioned protocols. 

\subsection{Datasets}
In each experiment related to the above mentioned attack scenarios, two datasets were generated: one training-set and one evaluation set. The training-set contained instances related to normal network behavior only, whereas the evaluation set contained instances related to normal and attacked periods (33.3\% labeled \emph{``normal''} and the remaining 66.67\% labeled \emph{``attack''}). \\

\subsection{Classifiers}
For evaluation purposes, we made use of three one-class algorithms: 
OC-PGA \cite{KontorovichHM11}, 1-SVM \cite{Scholkopf99estimatingthe}, and ADIFA (Attribute DIstribution Function Approximation) \cite{DBLP:journals/corr/abs-1209-1797}. We selected these base-classifiers because they represent the prominent families of one-class classifiers: density (OC-PGA) and boundary (OC-SVM). The OC-PGA algorithm is an adaptation for one-class learning from a well-known unsupervised algorithm \cite{Knorr97aunified} and the ADIFA algorithm is a univariate one-class anomaly detection algorithm based on meta-learning.
In training the meta-classifier (the classifier of the global anomaly detector), we made use of ADIFA because it performed considerably better than the other learning algorithms. 

\subsection{Performance Metrics}
\begin{figure}[ht]
	\centering
		\includegraphics[viewport= 100 260 750 570,scale =0.33]{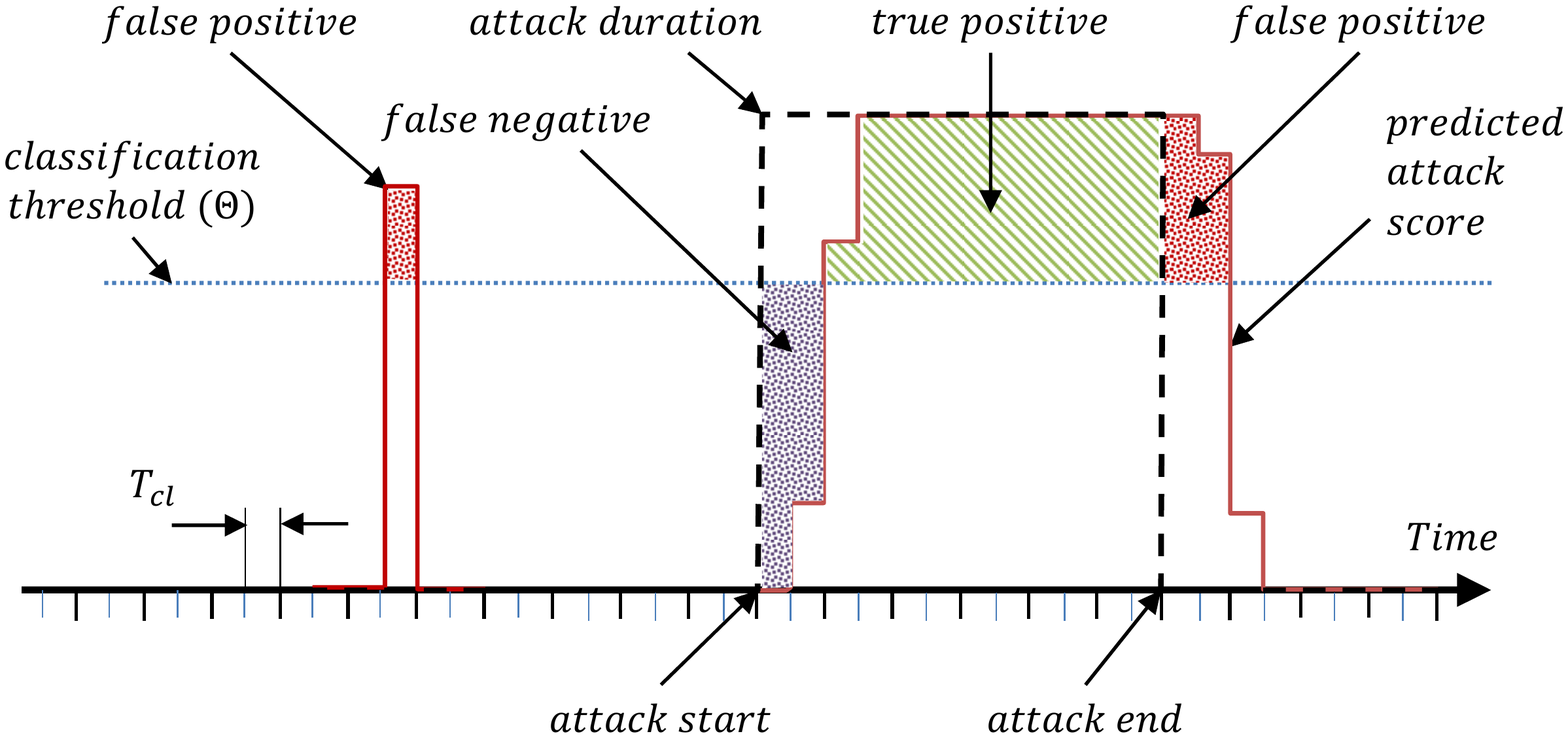} 
	\caption{A network attack surface and the corresponding predicted global network anomaly score over time}
	\label{fig:ClassificationPerformance}
\end{figure}

In evaluating the NIDS, the following performance metrics were used: classification error rate, F-Score, area under the ROC curve (AUC), and Time-to-Detect and anomaly-localization score. 
These performance metrics provide sufficient information to assess the fitness of the proposed NIDS for detecting network anomalies originated by network attacks. Figure \ref{fig:ClassificationPerformance} depicts a network attack that begins at ``$attack$ $start$'' time and ends after twelve $T_{cl}$ time-units. Also shown is the corresponding predicted attack score produced by the NIDS (i.e., $NAS$). 

The classification \emph{error} rate is the rate of incorrect predictions made by a classifier and is computed by the equation $error=(fp+fn)/(tp+tn+fp+fn)$, where $tp$, $tn$, $fp$ and $fn$ denote the number of true positive, true negative, false positive and false negative classifications of network behavior, respectively.

The \emph{F-Score} is a performance measure used in many research domains such as data mining and information retrieval. It is defined by $F$-$Score=2pr/(p+r)$, where $p$ and $r$ are classification precision and recall, respectively. To obtain a high score, both precision and recall must be high. Time-to-Detect (\emph{TtD}) is a temporal metric that measures the time (in time-units, TU) from the commencement of a network attack to detection by the classifier (the NIDS). Usually there is a trade-off between TtD and false-positive rate (FPR). A low TtD (a good measure) typically comes at the price of a greater FPR, since the classifier is less able to distinguish between noise signals, which are erratic, and attack signals, which tend to be more stable.

The \emph{Aread under the ROC curve (AUC)} measures the effectiveness of the classifier. The ROC curve is a graph produced by plotting the true positive rate $TPR$ (a.k.a $recall$) versus the false positive rate ($TPR=tp/(tp+fn), FPR=fp/(fp+tn)$). The AUC value of the best possible classifier will be equal to unity.
This would imply that it is possible to configure the classifier so that it will have 0\% false positive and 100\% true positive classifications. The worst possible binary classifier (obtained by flipping a coin for example) has AUC of 0.5.
AUC is considered as an objective performance metric as it does not depend on the specific discrimination threshold used by the classifier. 

Lastly, the anomaly-localization score (AL-Score) is an estimator that measures the extent to which the NIDS correctly identifies the source location of the anomaly. This measure is applicable for systems that can output a list of network components suspected of behaving anomalously. AL-Score is given by:
\[AL\-Score=1-\frac{\sum_{i=1}^{|An_{predicted}|}D(An_{predicted}^i,An_{real})}{\sum_{i=1}^{|v|}{D(v_{i},An_{real})}}\]
where $D(v_{i},U)$ computes the shortest path from node $v_i$ to the closest among nodes $U$; $An^i_{predicted}$ is the $i^{th}$ router in the group of routers predicted to be anomalous; and $An_{real}$ is the group of victim routers. 


\section{Experimental Results}
\label{sec:results}
In this section we present the empirical results, obtained by the proposed NIDS in the four of the above mentioned attack scenarios.\\

\subsection{Probes Traffic Volume} 
The probes are the most important tool the proposed NIDS uses for identifying network anomalies. Unfortunately, they inevitably add some traffic to the network. By doing so, they might affect the NIDS measurements, or even, if a significant volume of probe traffic is produced, can reduce the network QoS. Therefore, it is imperative to quantify the probes' effect on the network. Fortunately, since the probes dispatching scheme, probes packet size, and the re-sending frequency remain constant during the entire on-line phase, computing the augmented network traffic volume is straightforward. 

Table \ref{tab:ProbingTrafficProperties} summarizes the properties of the NIDS probes for the used network topologies. For calculating the probes traffic the following were assumed. A probe packet size is exactly 64 bytes, each probe is sent 25 times per second, and the bandwidth of every network link is 100Mb/s. The results indicate that the probes traffic volume is insignificant in terms of its capability to harm the network QoS.

\begin{table}[htbp]
	\centering
		\resizebox{1\linewidth}{!} {
		\begin{tabular}{llll} 
		\hline
		   & \#Probes 			& \multicolumn{2}{c}{Probe Traffic} \\ \cline{3-4}
		Topology & (Probes/Sec) & (Bytes/Sec) 	& \% of network's bandwidth\\
		\hline	
		1755 & \multicolumn{1}{c}{125}  & \multicolumn{1}{c}{8,000} & \multicolumn{1}{c}{0.004\%} 		\\
		3257 & \multicolumn{1}{c}{75}  & \multicolumn{1}{c}{4,800} & \multicolumn{1}{c}{0.005\%}		\\
		3356 & \multicolumn{1}{c}{1,375} & \multicolumn{1}{c}{88,000} & \multicolumn{1}{c}{0.007\%} 	\\
		4755 & \multicolumn{1}{c}{350} & \multicolumn{1}{c}{22,400} & \multicolumn{1}{c}{0.006\%}	\\
		\hline	
		\end{tabular}}
	\caption{Properties of the Probes traffic}
	\label{tab:ProbingTrafficProperties}
	\vspace{-2mm}
\end{table}
\vspace{-5mm}


\subsection{Comparison with Passive Features}
A basic assumption made in this study is that Probes' traffic features can yield a superior predictive model compared to passive-oriented feature sources. In this experiment we compared the Probes' features with two passive traffic features sources, i.e., link-based features and router-based features, with the intention of determining whether the primary assumption holds. In addition, we tested whether a mixture of feature sources can improve the NIDS classifiers' performance.
As link-based features, \emph{\#bytes/sec} and \emph{\#packet/sec} features were extracted for each link, and as router-based features, three management-information-base (MIB) features were extracted for each router: \emph{\#packets\_forwarded}, \emph{\#local\_deliveries} and \emph{\#unroutable\_packets}.

The NIDS was evaluated with both the partially disconnecting and the DNS cache poisoning attacks on the AS3356 and AS4755 network topology, respectively. We used only the ADIFA algorithm for training the Sensor classifiers.

\begin{figure}[t]
	\centering
		\includegraphics[scale=0.2]{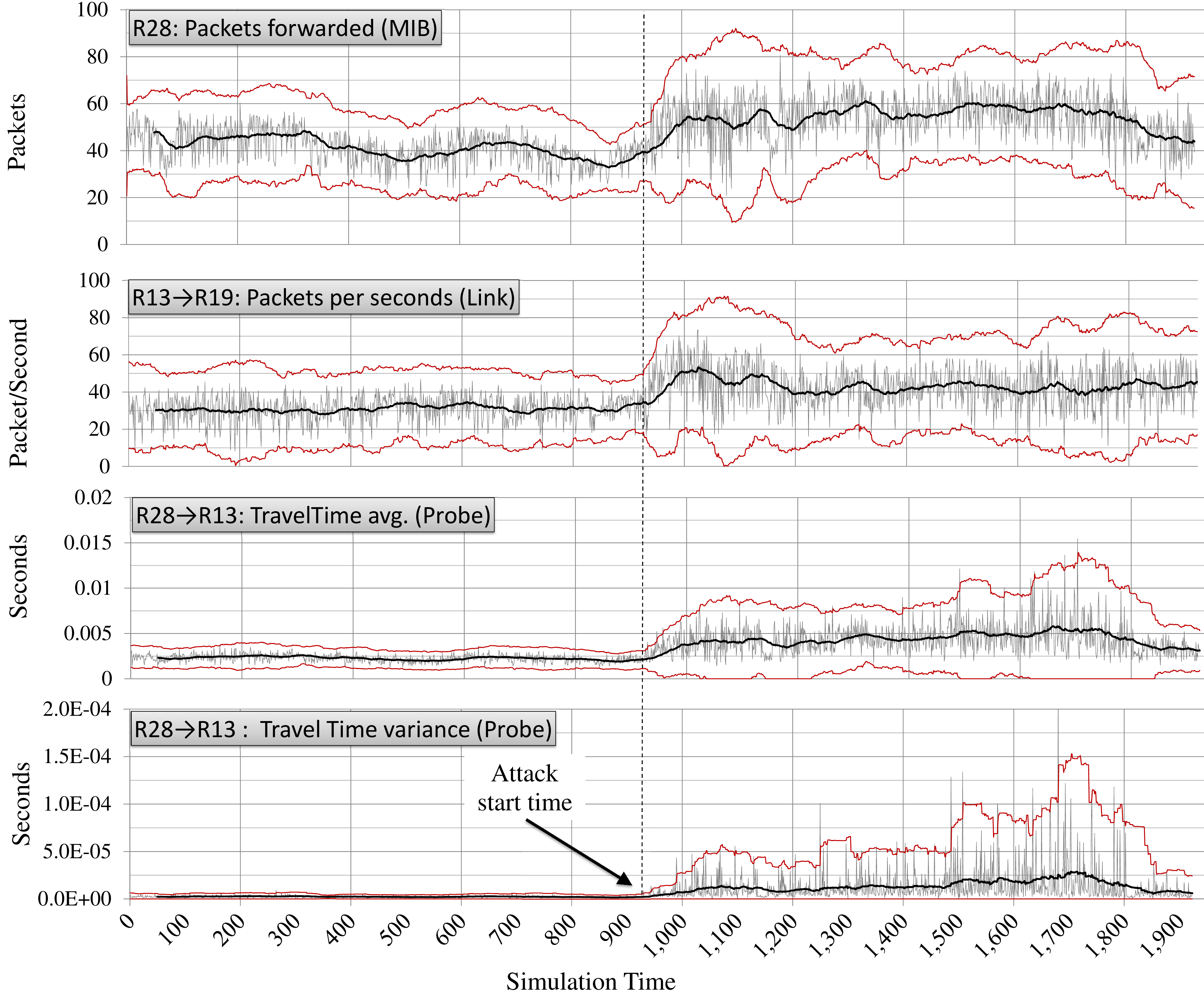} 
	\caption{Comparison between ``passive'' network features and ``active'' (probe) features}
	\label{fig:FeaturesPlot}
		\vspace{-3mm}
\end{figure}

Figure \ref{fig:FeaturesPlot} presents a stacked plot of four aggregate features (each data-point is an average of 25 raw values), extracted by the proposed NIDS during a simulation run, in which a DNS cache poisoning attack was executed after 925 seconds. 
The upper most graph shows the \emph{\#packets\_forwarded} feature, as measured in router R28, the second graph from the top shows the \emph{\#packets\_forwarded} feature of the R13 $\leftrightarrow$ R19 link, the third graph shows the R28 $\rightarrow$ R13 probe's round trip average \emph{travel time}, and the graph at the bottom shows the variance of the \emph{travel time} of the same probe. The data was captured at a simulation run, where a DNS cache poisoning attack was executed using the AS4755 topology. The gray graphs represent the raw data, the black bold graphs are the trend lines (using the moving average), and the red lines plot the three standard deviations range, computed from the raw value, over a historical time period of 60 seconds.

The graphs in Figure \ref{fig:FeaturesPlot} lead to two conclusions. First, prior to the attack, both the MIB and Link features are much less stable compared to the two probe features. Second, the mean and variance of the probes raw data are much more affected by the attack, compared to both the MIB and Link raw features; while the average and variance of the raw MIB values increased due to the attack by 27\% and 40\% respectively, the average and variance of the probe's raw data (travel time variance) were increased by 426\% and 1,680\% respectively.

\begin{table}[h]
	\centering
	\scriptsize
		\begin{tabular}{@{} p{20pt} @{} p{20pt} p{15pt} p{15pt} p{15pt} | p{15pt} p{15pt} p{15pt} p{15pt}} 
			\hline
			\begin{turn}{70}Metric \end{turn} & \begin{turn}{70}Victim Router\end{turn} 
							& \begin{turn}{70} \textbf{Probes}	\end{turn} & \begin{turn}{70} Links					 \end{turn} 
						  & \begin{turn}{70} MIBs							\end{turn} & \begin{turn}{70} Probe \& Links \end{turn} 
							&	\begin{turn}{70} Probe \& MIBs		\end{turn} & \begin{turn}{70} Links \& MIBs	 \end{turn}  
							& \begin{turn}{70} All Features			\end{turn} \\
			\hline
			\multirow{6}{*}{FPR}	
														 & R78 		& 0.000 & 0.000 & 0.000 & 0.000 & 0.000 & 0.000 & 0.000 \\
														 & R79 		& 0.000 & 0.096 & 0.153 & 0.086 & 0.143 & 0.146 & 0.138 \\
														 & R20 		& 0.000 & 0.118 & 0.124 & 0.103 & 0.135 & 0.126 & 0.153 \\
														 & R69 		& 0.000 & 0.167 & 0.205 & 0.156 & 0.185 & 0.224 & 0.223 \\
														 & R76 		& 0.000 & 0.175 & 0.141 & 0.060 & 0.132 & 0.150 & 0.146 \\ \cline{2-9}	
														 & Total 	& \textbf{0.000} & 0.111 & 0.124 & 0.081 & 0.119 & 0.129 & 0.132 \\
			\hline														
			\multirow{6}{*}{TPR}	
														 & R78 		& 0.990 & 0.968 & 0.722 & 0.988 & 0.989 & 0.901 & 0.988 \\
														 & R79 		& 0.989 & 0.664 & 0.989 & 0.989 & 0.988 & 0.988 & 0.986 \\
														 & R20 		& 0.885 & 0.998 & 0.990 & 0.994 & 0.990 & 0.990 & 0.990 \\
														 & R69 		& 0.329 & 0.787 & 0.342 & 0.723 & 0.346 & 0.606 & 0.591 \\
														 & R76 		& 0.989 & 0.993 & 0.990 & 0.991 & 0.990 & 0.988 & 0.989 \\ \cline{2-9}	
														 & Total 	& \textbf{0.836} & 0.882 & 0.806 & 0.937 & 0.861 & 0.894 & 0.909 \\
			\hline					
		\end{tabular}
	\caption{A comparison between ACTIDS features sources, based on TPR and FPR metrics} 
	\vspace{-3mm}
	\label{tab:FeaturesPower}
\end{table}

Continuing the comparison between the active and passive features, the next logical step is to study the classification performance of the proposed NIDS when using the different feature-set combinations. To this end, four scenarios of the partially-disconnecting attack were executed on the AS3356 topology, of which description is given in Table \ref{tab:PDAttackScenarios}.  

The results in Table \ref{tab:FeaturesPower} show that when trained using probe-based features, ACTIDS performed better compared to when trained on link-based or MIB-based features. A zero FPR was achieved only by using the probe-based features, while the other feature sources produced a considerably higher FPR. 

\begin{figure}[b]
	\centering
		\includegraphics[scale=0.38]{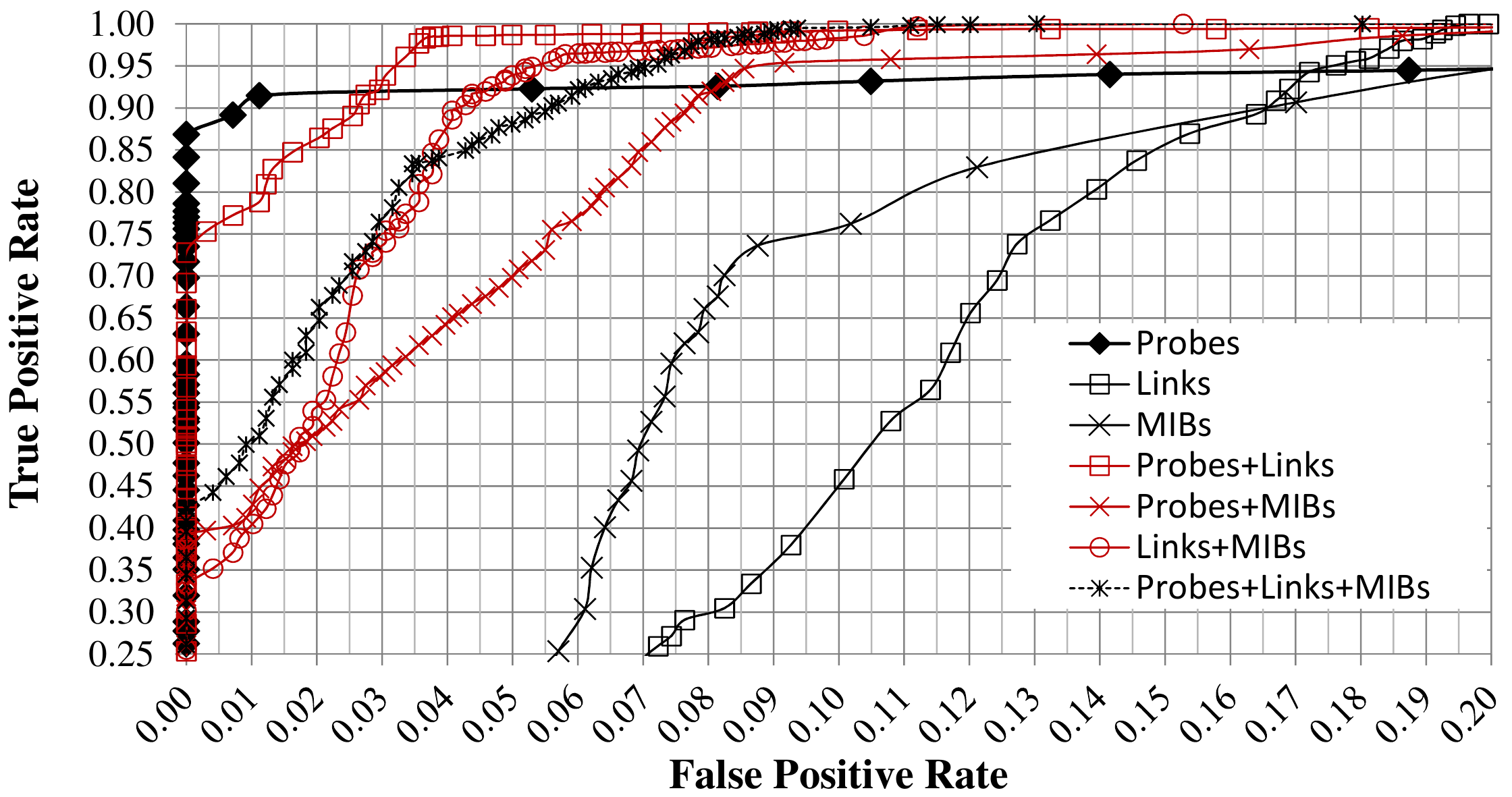} 
	\caption{ROC graphs for the DNS cache poisoning attack over the AS4755 topology}
	\label{fig:ROC-DNS-Poisoning}
	\vspace{-3mm}
\end{figure}

\begin{figure}[h]
	\centering
		\includegraphics[scale=0.38]{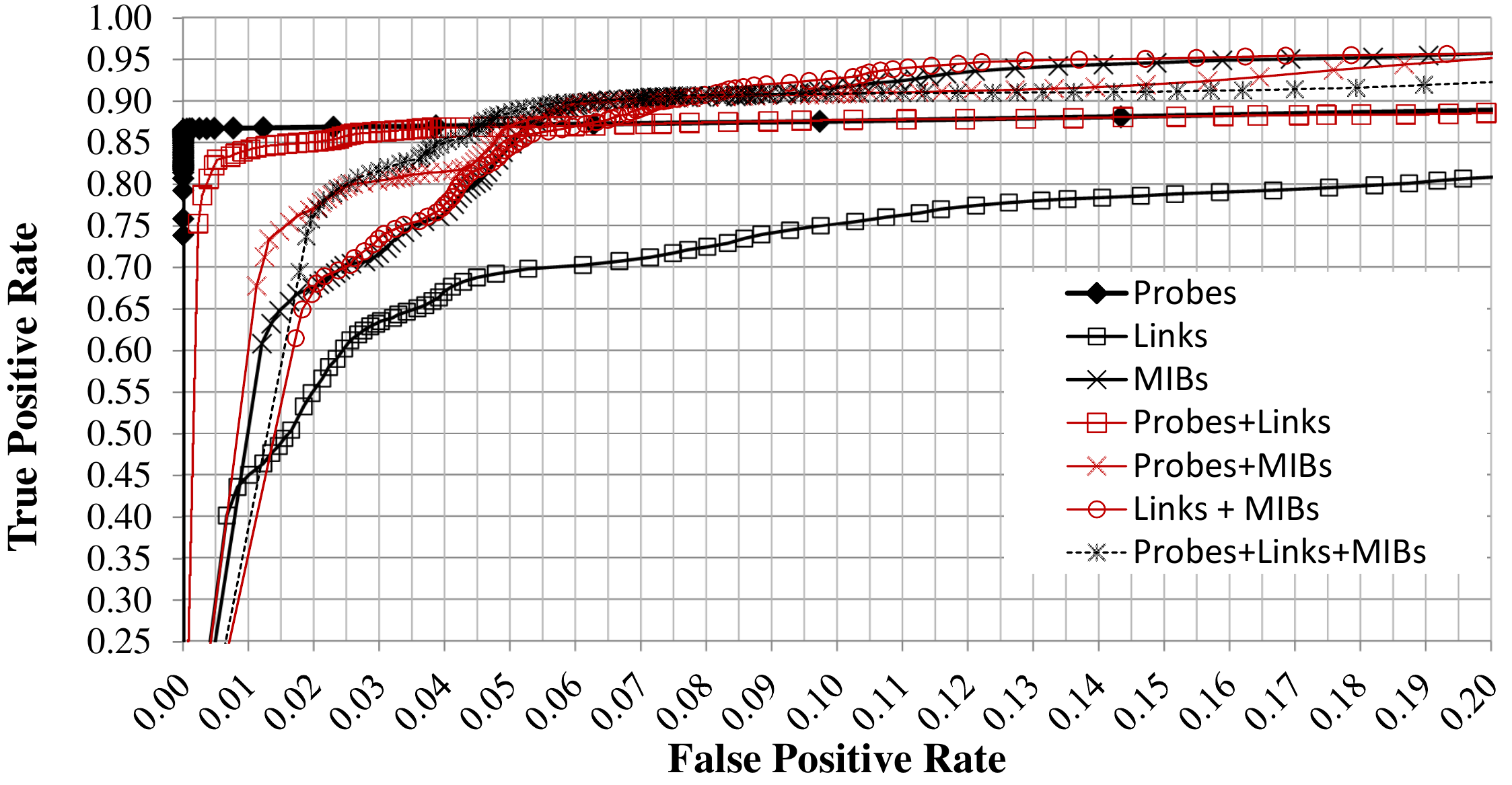} 
	\caption{ROC graphs for the partially-disconnecting attack over the AS3566 topology}
	\label{fig:ROC}
	\vspace{-3mm}
\end{figure}

In order to expand the generality of the above results, we plot ROC graphs by measuring the classification TPR and FPR of the proposed NIDS, for a range of classification thresholds. Figure \ref{fig:ROC-DNS-Poisoning} contains ROC graphs for the DNS cache poisoning attack, and Figure \ref{fig:ROC} contains ROC plots for the partially-disconnecting attack.
 
These ROC plots show that only the probe-feature group was able to produce a significant TPR (0.85 and 0.86 for the partially disconnecting and the DNS cache poisoning attacks, respectively) while, at the same time, inflicting no false positives at all.

 These results are highly significant because they suggest that probe-based NIDS can successfully detect zero-day attacks, without needing to monitor the attack's traffic or the attack objective, such as the OSPF routing table or the DNS cache. Interestingly, in both attack classes, the combination of probe features with either link or MIB features resulted with an inferior classification model, as long as zero FPR was required.

The following examines some additional aspects of the proposed NIDS. This examination is arranged in four sections, each of which deals with a different attack class.

\subsection{Detecting Partially Disconnecting Attacks}
\label{PartialDisconnectingAttacks}
In this experiment, ACTIDS is evaluated on attacks that logically disconnect half the victim(s) interfaces so that traffic can still pass or be routed through them on some interfaces. Hence the victim(s) are only partially ``disconnected''. This experiment has two goals: first, to study the ability of the presented NIDS to detect network-based attacks with global effects; second, to determine which machine-learning algorithm is best suited for detecting the abovementioned attacks. Table \ref{tab:PDAttackScenarios} summarizes the attack scenarios and the attributes of their corresponding datasets.

\begin{table}[htbp]
	\centering
		\scriptsize
		\begin{tabular} {|@{\hspace{0.5mm}}p{33pt}@{\hspace{0.5mm}}|@{\hspace{0.7mm}}p{108pt}@{\hspace{0.7mm}}||@{\hspace{0.7mm}}p{28pt}@{\hspace{0.7mm}}|@{\hspace{0.7mm}}p{50pt}|}
			\hline
			\multicolumn{2}{|c}{Attack Scenarios} & \multicolumn{2}{@{\hspace{-1.65mm}}||c|}{ACTIDS datasets} \\
			\hline
			Topology & \multicolumn{1}{c||@{\hspace{0.7mm}}}{Attacker $\rightarrow$ Victim} & Sensors & Features Avg.\\
			\hline
			AS3257 & R1$\rightarrow$ R3, R1$\rightarrow$ R6 															& \multicolumn{1}{c|@{\hspace{0.7mm}}}{4} 	& \multicolumn{1}{c@{}|}{12.5}  \\
			\hline
			AS1755 & R2$\rightarrow$ R11, R4$\rightarrow$ R13, R9$\rightarrow$ R15 					& \multicolumn{1}{c|@{\hspace{0.7mm}}}{4} 	& \multicolumn{1}{c@{}|}{22.5}  \\
			\hline
			AS4755 & R14$\rightarrow$ R13, R2$\rightarrow$ R11, R2$\rightarrow$ R16 				& \multicolumn{1}{c|@{\hspace{0.7mm}}}{15} 	& \multicolumn{1}{c@{}|}{12.06}  \\
			\hline
			\multirow{2}{*}{AS3356}	
				& R75$\rightarrow$ R78, R75$\rightarrow$ R79 & \multicolumn{1}{c|@{\hspace{0.7mm}}}{\multirow{2}{*}{54}} & \multicolumn{1}{c@{}|}{\multirow{2}{*}{10.1}}  \\
				& R77$\rightarrow$ R20, R8$\rightarrow$ R76 & &\\ 
			\hline
		\end{tabular}
	\caption{Partially Disconnecting attack scenarios} 
	\label{tab:PDAttackScenarios}
	\vspace{-3mm}
\end{table}

The results in Table \ref{tab:Experiment1Results} show that ADIFA algorithm achieve a detection rate greater than 98\% and that all three algorithms had no false-positive classifications: Their only classification errors were made during the attack period (false negative) when the NIDS was waiting to obtain ``sufficient'' global anomaly evidence.

\begin{table}[htb]
	\centering
	\scriptsize
		\begin{tabular}{@{}p{0.8cm}@{} @{}p{1.4cm}@{} 
										@{}p{1cm}  @{} @{}p{0.8cm}  @{} 
										@{}p{1cm}  @{} @{}p{1cm}  @{}
										cc  @{}} 
			\hline
		  \centering
		 AS & Algorithm & TPR & FPR & \% Error  & F-Score & TtD & AL-Score\\
			\hline
			\hline
			\multirow{4}{*}{3257} 	 	& 1-SVM		& 0.981 & 0.000 & 0.014 & 0.990 & 49.000 & 0.946 \\
																& OC-PGA 	& 0.980 & 0.000 & 0.014 & 0.990 & 50.500 & 0.945 \\
																& ADIFA  	& 0.982 & 0.000 & 0.013 & 0.991 & 46.500 & 0.942 \\
			\hline
			\multirow{4}{*}{1755}  	 	& 1-SVM  	& 0.981 & 0.000 & 0.014 & 0.990 & 49.000 & 0.951 \\
																& OC-PGA 	& 0.980 & 0.000 & 0.014 & 0.990 & 51.000 & 0.950 \\
																& ADIFA  	& 0.982 & 0.000 & 0.013 & 0.991 & 46.333 & 0.971 \\
			\hline
			\multirow{4}{*}{4755} 	 	& 1-SVM  	& 0.981 & 0.000 & 0.014 & 0.990 & 49.000 & 0.982 \\
																& OC-PGA 	& 0.681 & 0.000 & 0.229 & 0.710 & 49.667 & 0.980 \\
																& ADIFA  	& 0.982 & 0.000 & 0.013 & 0.991 & 46.333 & 0.993 \\
																 
			\hline
			\multirow{4}{*}{3356} 	 	& 1-SVM  	& 0.981 & 0.000 & 0.014 & 0.990 & 48.667 & 0.988 \\
																& OC-PGA 	& 0.793 & 0.000 & 0.149 & 0.856 & 48.333 & 0.986 \\
																& ADIFA  	& 0.982 & 0.000 & 0.013 & 0.991 & 47.000 & 0.998 \\
			\hline 
			\multirow{3}{*}{Total}  	& 1-SVM 	& 0.981 					& 0.000 & 0.014 & 0.990 & 48.909 					& 0.969 \\
																& OC-PGE  & 0.847 					& 0.000 & 0.110 & 0.877 & 49.818 					& 0.967 \\
																& ADIFA  	& \textbf{0.982} 	& 0.000 & 0.013 & 0.991 & \textbf{46.474} & \textbf{0.976} \\
			\hline \hline
			\multirow{2}{*}{Total}		&	$EP$  	& 0.878 					& 0.007 & 0.090 & 0.886 & \textbf{28.810} & \multirow {2}{*}{0.970} \\	\cline{2-7}
																&	$NAS$   & \textbf{0.934} 	& 0.000 & 0.047 & 0.951 & 48.492 &  \\
		\hline
		\end{tabular}
	\caption{Result table for the partially disconnecting attack class}
	\vspace{-7mm}
	\label{tab:Experiment1Results}
\end{table} 

Comparison of the two presented output functions reveals that EP (ensemble prediction output) is the most responsive measure for detecting attacks; however, it comes at the price of a higher FPR. In contrast, $NAS$ has longer TtD than $EP$, but generates fewer false-positive classifications, which is usually a more important property for real NIDS.

\subsection{Detecting Link Weight Distortion Attacks}
\label{SubversionAttacks}
Continuing further with the experimentation, we evaluate the proposed NIDS on the `Link Weight Distortion' attack over three different scenarios. In the first two scenarios a single link weight is affected, whereas in the third, two links are simultaneously affected. Specifically, the implemented attack increased the target link's weight (cost) by one orders of magnitude, to insure a global anomaly. The experiment was performed on the largest network topology, i.e., AS3356. 

\begin{table}[htb]
	\centering
	\scriptsize
		\begin{tabular}{@{}p{1.8cm}@{} @{}p{1.3cm}@{} 
										@{}p{0.8cm}  @{} @{}p{0.8cm}  @{} 
										@{}p{1cm}  @{} @{}p{1.0cm}  @{}
										@{}p{0.6cm}  @{} @{}p{1.2cm}  @{}} 
			\hline
		 Affected Link & Algorithm & TPR & FPR & \%Error  & F-Score & TtD & AL-Score\\
			\hline
			\hline
				\multirow{3}{*}{R50$\leftrightarrow$ R79}  			
																			& 1-SVM 	& 0.000 					& 0.000 & 0.667 & 0.0 	& n/a 	& \multicolumn{1}{c}{\multirow{1}{*}{n/a}} \\
																			& OC-PGA  & 0.913 					& 0.000 & 0.062 & 0.955 & 33.0 	& \multicolumn{1}{c}{\multirow{1}{*}{0.984}} \\
																			& ADIFA  	& \textbf{0.985} 	& 0.000 & 0.011 & 0.992 & 18.0	& \multicolumn{1}{c}{\multirow{1}{*}{0.993}} \\

			\hline
				\multirow{3}{*}{R79$\leftrightarrow$ R80} 				
																			& 1-SVM  	& 0.000 & 0.000 & 0.667 & 0.0 	& n/a		& \multicolumn{1}{c}{\multirow{1}{*}{n/a}} \\
																			& OC-PGA  & 0.000 & 0.000 & 0.667 & 0.0 	& n/a		& \multicolumn{1}{c}{\multirow{1}{*}{n/a}} \\
																			& ADIFA  	& 0.468 & 0.000 & 0.379 & 0.637 & 136.0 & \multicolumn{1}{c}{\multirow{1}{*}{0.994}} \\
			\hline \hline
							\multirow{3}{*}{\vbox{R50$\leftrightarrow$ R79 R79$\leftrightarrow$ R80}} 				
																			& 1-SVM 	& 0.000 					& 0.000 & 0.667 & 0.0 	& n/a 	& \multicolumn{1}{c}{\multirow{1}{*}{n/a}} \\
																			& OC-PGA  & 0.990 					& 0.000 & 0.667 & 0.995 & 25.0 	& \multicolumn{1}{c}{\multirow{1}{*}{0.984}} \\
																			& ADIFA		& \textbf{0.995} 	& 0.000 & 0.003 & 0.998 & 13.0 	& \multicolumn{1}{c}{\multirow{1}{*}{0.993}} \\
		\hline
		\end{tabular}
	\caption{Result table for the Link-Weight Distortion attack class} 
	\vspace{-4mm}
	\label{tab:Experiment2Results}
\end{table} 

The results in Table \ref{tab:Experiment2Results} show that the Link Weight Distortion attack was, in general, harder to detect, when compared to the partially disconnecting attack. Consequently, when the proposed NIDS used the OC-PGA classifier, it did not detect one of the three attacks, and with the 1-SVM classifier it missed the attacks of all three scenarios. In contrast, when it used the ADIFA classifiers it detected all the attacks, and presented the best classification performance among the examined classifiers, with the shortest time-to-detect and the highest true positive detection rate.

\subsection{Detecting DNS Cache Poisoning Attack}
In this section we study the detection performance of the proposed NIDS on a simulated effect of the DNS cache poisoning. Specifically, we examine the NIDS sensitivity using an increasing attack intensity, implemented by a series of experiments, each simulates an increased percentage of DNS poisoned entries at the victim routers. Five DNSs (routers) were subjected attacked during the experiment series to insure a substantial effect on the network's QoS, even when only 10 percents of the DNS cache entries are poisoned. The experiment series was performed using the AS4755 topology. Router $R28$ was used as the adversary, whereas the victims role was given to routers $R1$, $R2$, $R3$, $R8$, and $R9$.

\begin{figure}[h]
	\centering
		\includegraphics[scale=0.51]{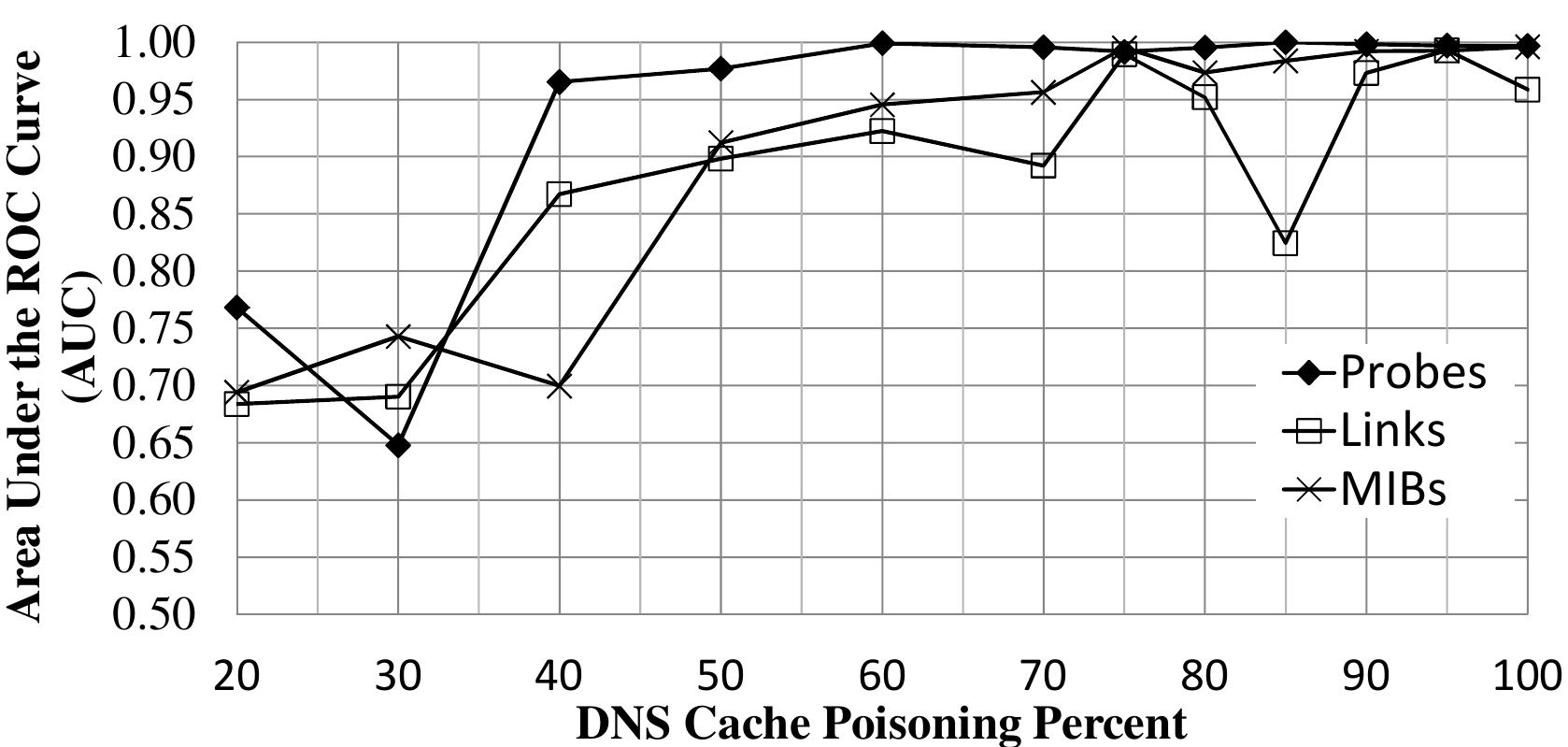} 
	\caption{AUC results for the proposed NIDS over different DNS cache poisoning percentage}
	\label{fig:NDSPoisoningPercent}
	\vspace{-1mm}
\end{figure}

The results in Figure \ref{fig:NDSPoisoningPercent} indicates that when the cache poisoning attack intensity was below 40 percent, non of the feature sets provided sufficient and robust information for intrusion detection. A change of trends started from an intensity around 40 to 50 percent. From this point and on, the Probe features incurred a detection AUC rate of above 0.95, which was greater than that of both the Link and MIB feature sets. It is important to note that until the attack reached the critical value of around 40 percents, the network QoS remained unharmed, as no evidence of increased delays, jitter or packet loss could be observed, either automatically or manually, from the extracted raw network data.

\subsection{Detecting Authoritative DNS Server Poisoning Attacks}
\begin{figure}[b]
	\centering
		\includegraphics[scale=0.51]{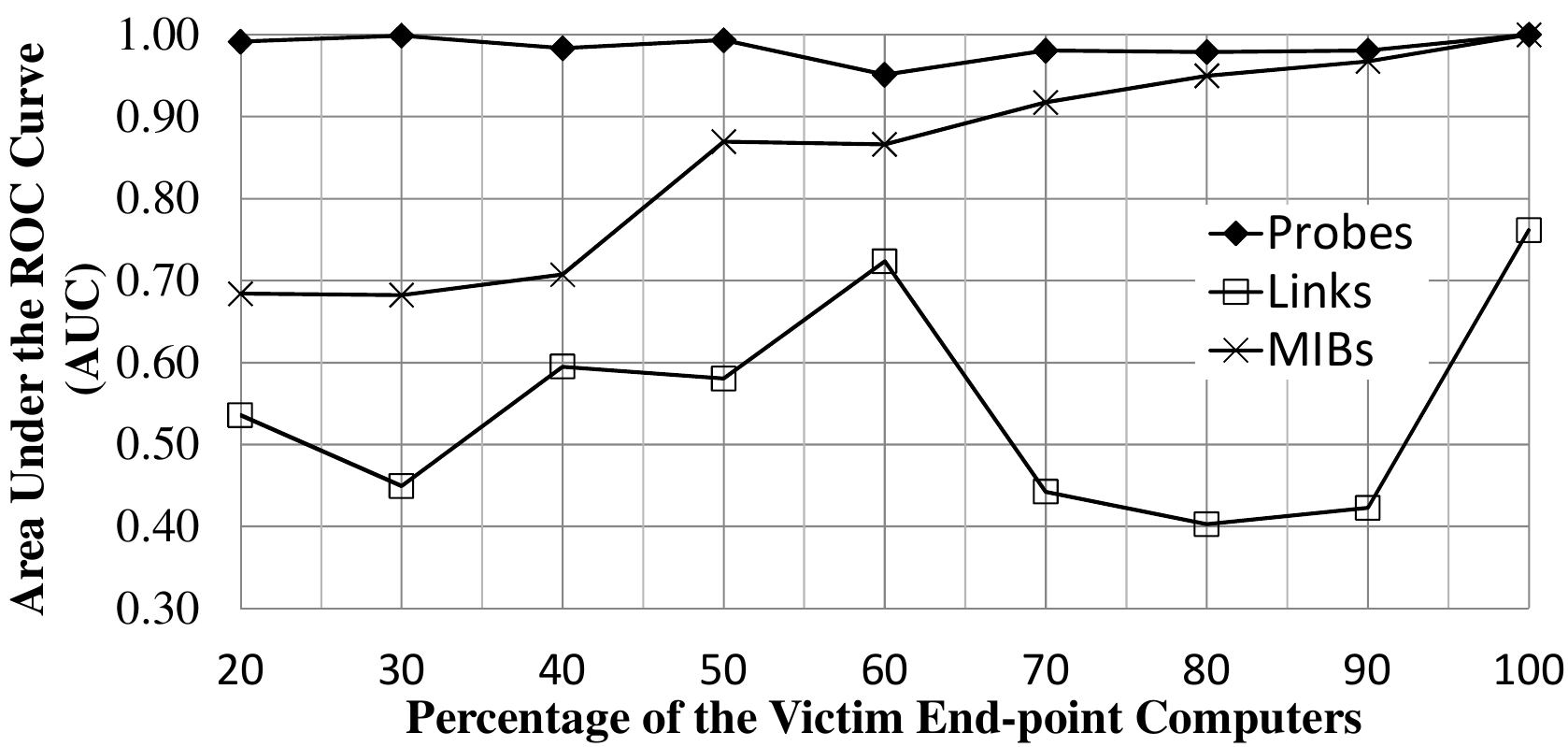} 
	\caption{AUC results for the proposed NIDS over different DNS cache poisoning percentage}
	\label{fig:DNS-Authoritative-Percent}
	\vspace{-3mm}
\end{figure}

The current experiment studies the robustness of the proposed NIDS in detecting the authoritative DNS server poisoning attack.
Unlike the DNS cache poisoning attack, every endpoint computer is subjected to the effect of the attack. Therefore, in order to generate a moderate, yet significant attack effect, only a single DNS entry was poisoned, so that only a small fraction of the global network traffic would be affected. In order to control the attack's  intensity, we limited the number of endpoints computers in the local network behind the victim router that are subjected to the attack. For example, if the local network contains 10 endpoint computers, and the attack intensity is 20 percents, only the  traffic directed to a selected 2 endpoint-computers behind the victim router will be redirected to the adversary. The experiment was conducted using the AS3257 topology, of which results are presented in Figure \ref{fig:DNS-Authoritative-Percent}.

The results show that the QoS of the network had deteriorated, even when only 20 percents of the endpoint computers behind the victim router were affected. The greatest efficiency in the intrusion detection task was achieved by using the Probe features. In contrast, the Link features provided the least useful information among the three available feature sets. The MIB features provided a detection performance, similar to that achieved by the Probe features, only when the attack intensity was greater than 80 percents.

\subsection{Handling Byzantine Faults}
\label{sec:LinkFailures}
Byzantine Faults are benign network events in which some network elements perform significantly differently to that described by their nominal definition due to common reasons such as failed links, routers, switches or other network hardware. Naturally, these types of devices are monitored by the network service provider's dedicated control channels, and therefore, detecting such events is not the focus of the present work. However, because these benign events might incur false detections for ACTIDS, we would like to study its ability to discriminate benign anomalies from anomalies that are the result of malicious activity.

The next experiment examines the extent to which ACTIDS detects benign anomalies and how it is trained to ignore such anomalies.
To simulate benign routing anomalies, we incurred some link-failures during the simulation execution. To make the benign anomalies more interesting, the link-failures were designed to disconnect some critical links, i.e, non-redundant links between central routers, so that when executed the network would be strongly affected. Two links were chosen: $R76 \leftrightarrow R80$ and $R51 \leftrightarrow R80$.
The links were disconnected at simulation time $TU_{start}$=182, 2626, and then were returned to full connectivity at $TU_{end}$=1980 and 3480 respectively.

In this experiment, performed over the AS3356 network, the NIDS' classifiers were trained by the ADIFA algorithm on three sets of features: Probes (denoted $P$), Links (denoted $L$) and Probes + Links (denoted $PL$). $Tr$ denotes a training-set containing normal network behavior instances, $Tr_{+}$ denotes training-sets that contain both normal and benign anomaly instances, and $valid. set$ represents a validation set incorporating both normal instances and attack (partially disconnecting) instances. 

\begin{table}[htpb]
	\centering
	\resizebox{1\linewidth}{!} {
		\begin{tabular}{@{} p{23pt} p{28pt} c| c c |c c|c c| c @{}} 
			\hline
			\multirow{2}{*}{Training} & \multirow{2}{*}{Validation} & 	\multirow{2}{*}{Feature} &
			\multicolumn{2}{c|}{FPR} & \multicolumn{2}{c|}{TPR} & \multicolumn{2}{c|}{FNR} & \multirow{2}{*}{TtD}\\
														 &	&	& $EP$ & $NAS$ & $EP$ & $NAS$ & $EP$ & $NAS$ & \\
			\hline
			  \multirow{3}{*}{$Tr$} & \multirow{3}{*}{$Tr_{+}$} 
																										& $P$		& 0.02 & 0 		& $-$ & $-$ & $-$ & $-$ & $-$ \\
																								&  	&	$L$	 	& 0.26 & 0.19 & $-$ & $-$ & $-$ & $-$ & $-$ \\
																								&	 	&	$PL$ 	& 0.72 & 0.67 & $-$ & $-$ & $-$ & $-$ & $-$ \\ 

			\hline 
			\multirow{3}{*}{$Tr_{+}$} & \multirow{3}{*}{$Tr_{+}$}	
																										& $P$		& 0.01 & 0 		& $-$ & $-$ & $-$ & $-$ & $-$ \\
																								&	 	& $L$ 	& 0.22 & 0.11 & $-$ & $-$ & $-$ & $-$ & $-$ \\
																								&	 	& $PL$ 	& 0.09 & 0.07 & $-$ & $-$ & $-$ & $-$ & $-$ \\ 
			\hline
			\multirow{3}{*}{$Tr_{+}$} & \multirow{3}{*}{$valid.set$} 
																										& $P$		& 0.02 & 0	& 1 		& 0.99 & 0.00 & 0.01 & 28.5	\\
																								&		& $L$		& 0.06 & 0 	& 0.64 	& 0.57 & 0.34 & 0.43 & 291.6 \\
																								&		& $PL$	& 0.06 & 0 	& 1 	 	& 0.99 & 0.00 & 0.01 & 25.3 	\\
			\hline
			
		\end{tabular}}	
	\caption{The effect of Byzantine Faults on the NIDS classification performance}

	\label{tab:LinkFailure1Results}
		\vspace{-6mm}
\end{table} 

Table \ref{tab:LinkFailure1Results} shows that benign anomalies can indeed be falsely identified as network attacks and that the choice of training features has a large impact on the false detection rate. When trained with \emph{Probes} features, ACTIDS had a 2\% false detection rate; however, when trained with \emph{Link} features, it had an imposing 26\% false detection rate.

Continuing with the experiment, we explored how to mitigate the false detection. We believe that, by modeling benign anomalies as part of the normal state, ACTIDS will learn to ignore similar benign anomalies. For this reason, ACTIDS was trained upon both normal instances and link-failure instances. To evaluate the effect of these new conditions, ACTIDS was evaluated on normal and attack instances. The results are presented in the final three rows of Table \ref{tab:LinkFailure1Results}. It appears that, in these new conditions, ACTIDS made much fewer detection errors, regardless of the feature-set used. Surprisingly, the results indicate that modeling benign anomalies, instead of rendering the NIDS less susceptible to any type of anomaly, did not harm the performance in any measurable way; the $recall$ remained higher than 0.98 while maintaining a zero FPR and the $TtD$ was not prolonged as expected.

\section{Summary and Conclusions}
\label{sec:conclusions}
In this paper we have presented a new active strategy for NIDS called ACTIDS. It detects and localizes network attacks by means of self-produced short-range probes, hierarchical anomaly detection and one-class ensemble learning. 

We examined the proposed NIDS on 4 network topologies over 35 attack scenarios. The smallest network contained 9 routers, whereas the largest contained 82 routers. The experiments spanned five dimensions: attack class, features groups, topology, learning algorithm, and the NIDS output function. 
Our results show that the active approach can perform better than the passive approach in detecting attacks that harm the QoS of the network. Experimented on the same platform, the active Probes features produced a considerably better classification performance, compared to the passive features. In particular, the Probe features incurred a zero false-detection rate (FPR), while achieving a true-detection rate ($recall$) of 85\%. In order to produce such a high recall value, the NIDS, when trained on passive-related features, suffered from more than 10\% false-detection classifications.

Another strength of the proposed NIDS is with its anomaly-localizing technique, which has been shown to be highly accurate and therefore, can serve as a powerful forensic tool for network operators and security experts.

The time taken for ACTIDS to detect a network attack was, on average, 46.8 seconds. This is an excellent result bearing in mind the zero-false-positives constraint and that network attacks, within the scope of this paper, are expected to last longer than an order of minutes.


\bibliographystyle{IEEEtran} 
\bibliography{mybib}

\begin{thebibliography}{10}
\providecommand{\url}[1]{#1}
\csname url@samestyle\endcsname
\providecommand{\newblock}{\relax}
\providecommand{\bibinfo}[2]{#2}
\providecommand{\BIBentrySTDinterwordspacing}{\spaceskip=0pt\relax}
\providecommand{\BIBentryALTinterwordstretchfactor}{4}
\providecommand{\BIBentryALTinterwordspacing}{\spaceskip=\fontdimen2\font plus
\BIBentryALTinterwordstretchfactor\fontdimen3\font minus
  \fontdimen4\font\relax}
\providecommand{\BIBforeignlanguage}[2]{{%
\expandafter\ifx\csname l@#1\endcsname\relax
\typeout{** WARNING: IEEEtran.bst: No hyphenation pattern has been}%
\typeout{** loaded for the language `#1'. Using the pattern for}%
\typeout{** the default language instead.}%
\else
\language=\csname l@#1\endcsname
\fi
#2}}
\providecommand{\BIBdecl}{\relax}
\BIBdecl

\bibitem{PaxonSommer10}
R.~Sommer and V.~Paxson, ``Outside the closed world: On using machine learning
  for network intrusion detection,'' in \emph{Security and Privacy (SP), 2010
  IEEE Symposium on}, may 2010, pp. 305 --316.

\bibitem{Lakhina05}
A.~Lakhina, M.~Crovella, and C.~Diot, ``Mining anomalies using traffic feature
  distributions,'' in \emph{Proceedings of SIGCOMM}, 2005, pp. 217--228.

\bibitem{Brauckhoff06}
D.~Brauckhoff, B.~Tellenbach, A.~Wagner, M.~May, and A.~Lakhina, ``Impact of
  packet sampling on anomaly detection metrics,'' in \emph{Proceedings of ACM
  SIGCOMM on Internet measurement}, 2006, pp. 159--164.

\bibitem{weka}
I.~H. Witten and E.~Frank, \emph{Data Mining: Practical machine learning tools
  and techniques}, 2nd~ed.\hskip 1em plus 0.5em minus 0.4em\relax San
  Francisco: Morgan Kaufmann, 2005.

\bibitem{Barford09}
P.~Barford, N.~Duffield, A.~Ron, and J.~Sommers, ``Network performance anomaly
  detection and localization,'' in \emph{INFOCOM 2009, IEEE}, april 2009, pp.
  1377 --1385.

\bibitem{Clark10}
D.~D. Clark and S.~Landau, ``Untangling attribution,'' in \emph{Workshop on
  Deterring Cyberattacks: Informing Strategies and Developing Options for U.S.
  Policy}, 2010.

\bibitem{Schapire90}
R.~E. Schapire, ``The strength of weak learnability,'' \emph{Machine Learning},
  vol.~5, pp. 197--227, 1990.

\bibitem{MenahemRE09}
E.~Menahem, L.~Rokach, and Y.~Elovici, ``Troika - an improved stacking schema
  for classification tasks,'' \emph{Inf. Sci.}, vol. 179, no.~24, pp.
  4097--4122, 2009.

\bibitem{GiacintoPRR08}
G.~Giacinto, R.~Perdisci, M.~D. Rio, and F.~Roli, ``Intrusion detection in
  computer networks by a modular ensemble of one-class classifiers,''
  \emph{Information Fusion}, vol.~9, no.~1, pp. 69--82, 2008.

\bibitem{Tax01}
D.~M. Tax and R.~P. Duin, ``Combining one-class classifiers,'' in \emph{in
  Proc. Multiple Classifier Systems, 2001}.\hskip 1em plus 0.5em minus
  0.4em\relax Springer Verlag, 2001, pp. 299--308.

\bibitem{JuszczakD04}
P.~Juszczak and R.~P.~W. Duin, ``Combining one-class classifiers to classify
  missing data,'' in \emph{Multiple Classifier Systems}, 2004, pp. 92--101.

\bibitem{DBLP:journals/corr/abs-1112-5246}
E.~Menahem, L.~Rokach, and Y.~Elovici, ``Combining one-class classifiers via
  meta-learning,'' \emph{CoRR}, vol. abs/1112.5246, 2011.

\bibitem{Lee01realtime}
W.~Lee, S.~J. Stolfo, P.~K. Chan, E.~Eskin, W.~Fan, M.~Miller, S.~Hershkop, and
  J.~Zhang, ``Real time data mining-based intrusion detection,'' 2001.

\bibitem{SalemVG10}
O.~Salem, S.~Vaton, and A.~Gravey, ``A scalable, efficient and informative
  approach for anomaly-based intrusion detection systems: theory and
  practice,'' \emph{Int. Journal of Network Management}, vol.~20, no.~5, pp.
  271--293, 2010.

\bibitem{DBLP:journals/tdsc/GuptaNR10}
K.~K. Gupta, B.~Nath, and K.~Ramamohanarao, ``Layered approach using
  conditional random fields for intrusion detection,'' \emph{IEEE Trans.
  Dependable Sec. Comput.}, vol.~7, no.~1, pp. 35--49, 2010.

\bibitem{DBLP:journals/tdsc/WangZS04}
H.~Wang, D.~Zhang, and K.~G. Shin, ``Change-point monitoring for the detection
  of dos attacks,'' \emph{IEEE Trans. Dependable Sec. Comput.}, vol.~1, no.~4,
  pp. 193--208, 2004.

\bibitem{DBLP:journals/tdsc/HwangCCQ07}
K.~Hwang, M.~Cai, Y.~Chen, and M.~Qin, ``Hybrid intrusion detection with
  weighted signature generation over anomalous internet episodes,'' \emph{IEEE
  Trans. Dependable Sec. Comput.}, vol.~4, no.~1, pp. 41--55, 2007.

\bibitem{HubballiBN10}
N.~Hubballi, S.~Biswas, and S.~Nandi, ``Layered higher order n-grams for
  hardening payload based anomaly intrusion detection,'' in \emph{ARES}, 2010,
  pp. 321--326.

\bibitem{Sommer03}
R.~Sommer, ``Bro: An open source network intrusion detection system,'' in
  \emph{DFN-Arbeitstagung {\"u}ber Kommunikationsnetze}, 2003, pp. 273--288.

\bibitem{OlivainG05}
J.~Olivain and J.~Goubault-Larrecq, ``The orchids intrusion detection tool,''
  in \emph{CAV}, 2005, pp. 286--290.

\bibitem{BenaliBGC10}
F.~Benali, N.~Bennani, G.~Gianini, and S.~Cimato, ``A distributed and
  privacy-preserving method for network intrusion detection,'' in \emph{OTM
  Conferences (2)}, 2010, pp. 861--875.

\bibitem{Bao09}
C.-M. Bao, ``Intrusion detection based on one-class svm and snmp mib data,'' in
  \emph{IAS}, 2009, pp. 346--349.

\bibitem{Jou00designand}
Y.~F. Jou, F.~Gong, J.~G. Sargor, X.~Wu, S.~F. Wu, H.~C. Chang, and F.~Wang,
  ``Design and implementation of a scalable intrusion detection system for the
  protection of network infrastructure,'' in \emph{In DARPA Information
  Survivability Conference and Exposition}, 2000, pp. 25--27.

\bibitem{NassarSF10}
M.~Nassar, R.~State, and O.~Festor, ``A framework for monitoring sip enterprise
  networks,'' in \emph{NSS}, 2010, pp. 1--8.

\bibitem{DBLP:journals/ton/BejeranoR06}
Y.~Bejerano and R.~Rastogi, ``Robust monitoring of link delays and faults in ip
  networks,'' \emph{IEEE/ACM Trans. Netw.}, vol.~14, no.~5, pp. 1092--1103,
  2006.

\bibitem{DBLP:conf/infocom/ChengQMQB10}
L.~Cheng, X.~Qiu, L.~Meng, Y.~Qiao, and R.~Boutaba, ``Efficient active probing
  for fault diagnosis in large scale and noisy networks,'' in \emph{INFOCOM},
  2010, pp. 2169--2177.

\bibitem{DYSWIS08}
V.~Singh, H.~Schulzrinne, and K.~Miao, ``Dyswis: An architecture for automated
  diagnosis of networks,'' in \emph{Network Operations and Management
  Symposium, 2008. NOMS 2008. IEEE}.\hskip 1em plus 0.5em minus 0.4em\relax
  IEEE, 2008, pp. 851--854.

\bibitem{BarbhuiyaBN11}
F.~A. Barbhuiya, S.~Biswas, and S.~Nandi, ``An active des based ids for arp
  spoofing,'' in \emph{SMC}, 2011, pp. 2743--2748.

\bibitem{SalhiLC10}
E.~Salhi, S.~Lahoud, and B.~Cousin, ``Joint optimization of monitor location
  and network anomaly detection,'' in \emph{LCN}, 2010, pp. 204--207.

\bibitem{VargaH08}
A.~Varga and R.~Hornig, ``An overview of the omnet++ simulation environment,''
  in \emph{SimuTools}, 2008, p.~60.

\bibitem{SpringMW02}
N.~T. Spring, R.~Mahajan, and D.~Wetherall, ``Measuring isp topologies with
  rocketfuel,'' in \emph{SIGCOMM}, 2002, pp. 133--145.

\bibitem{Kowalski95modellingtraffic}
J.~P. Kowalski and B.~Warfield, ``Modelling traffic demand between nodes in a
  telecommunications network,'' in \emph{in ATNAC}, 1995.

\bibitem{self-similar}
W.~Leland, M.~Taqqu, W.~Willinger, and D.~Wilson, ``On the self-similar nature
  of ethernet traffic (extended version),'' \emph{Networking, IEEE/ACM
  Transactions on}, vol.~2, no.~1, pp. 1 --15, feb 1994.

\bibitem{DiversionAttacks}
E.~Menahem, G.~Nakibly, and Y.~Elovici, ``{Degrading the Network's Quality of
  Service via Traffic Diversion Attacks},''
  \url{https://www.dropbox.com/s/tcz4g03xq844fae/TDA.pdf}, Tech. Rep., 12 2012.

\bibitem{KontorovichHM11}
A.~Kontorovich, D.~Hendler, and E.~Menahem, ``Metric anomaly detection via
  asymmetric risk minimization,'' in \emph{SIMBAD}, 2011, pp. 17--30.

\bibitem{Scholkopf99estimatingthe}
B.~Sch\"{o}lkopf, J.~C. Platt, J.~Shawe-taylor, A.~J. Smola, and R.~C.
  Williamson, ``Estimating the support of a high-dimensional distribution,''
  1999.

\bibitem{DBLP:journals/corr/abs-1209-1797}
\BIBentryALTinterwordspacing
E.~Menahem, A.~Schclar, L.~Rokach, and Y.~Elovici, ``Securing your
  transactions: Detecting anomalous patterns in xml documents,'' \emph{CoRR},
  vol. abs/1209.1797, 2012. [Online]. Available:
  \url{http://arxiv.org/abs/1209.1797}
\BIBentrySTDinterwordspacing

\bibitem{Knorr97aunified}
E.~M. Knorr and R.~T. Ng, ``A unified notion of outliers: Properties and
  computation,'' in \emph{In Proc. of the International Conference on Knowledge
  Discovery and Data Mining}.\hskip 1em plus 0.5em minus 0.4em\relax AAAI
  Press, 1997, pp. 219--222.

\end{thebibliography}


\end{document}